\documentclass[twocolumn,tighten,10pt]{aastex631}
\usepackage{appendix}
\usepackage{graphicx}
\usepackage{tablefootnote}
\usepackage{longtable}

\newcommand{\degree}{^\circ}

\usepackage{xcolor}

\begin{document}

\title{Establishing a Connection Between the Jet and the Corona in Black Hole Low Mass X-Ray Binaries }

\author[0009-0009-3384-2830]{Eric M. Davidson}
\affiliation{Department of Physics, Montana State University, P.O. Box 173840, Bozeman, MT 59717-3840, USA}
\author[0000-0003-0330-1901]{Jaiverdhan Chauhan}
\affiliation{School of Physics and Astronomy, University of Leicester, University Road, Leicester LE1 7RH, UK}
\author[0009-0000-9468-7277]{Anne Lohfink}
\affiliation{Department of Physics, Montana State University, P.O. Box 173840, Bozeman, MT 59717-3840, USA}
\author[0000-0002-7930-2276]{Thomas D. Russell}
\affiliation{INAF/IASF Palermo, via Ugo La Malfa 153, I-90146 Palermo, Italy}
\author[0000-0001-5803-2038]{Rhaana Starling}
\affiliation{School of Physics and Astronomy, University of Leicester, University Road, Leicester LE1 7RH, UK}
\author{Charlotte Johnson}
\affiliation{Department of Physics \& Astronomy, University of South Carolina, Columbia, SC 29208, USA}

\begin{abstract}

Low-mass black hole X-ray Binaries (LMXBs) undergo outbursts, during which their brightness increases greatly for timescales of months. The X-ray accretion and radio jet properties change dramatically throughout an outburst in a broadly consistent way between sources. Changes to the accretion flow and the corona are evident through X-ray spectral variations, while the jet's evolution produces changes in the radio. Typically, high energy emission from the corona initially dominate the X-ray spectrum, and quasi-steady compact jets are observed in the radio. As the outburst progresses, emission from the corona fades and is superseded by lower energy X-ray accretion disk emission. During this transition, the compact jets are quenched and discrete ejecta, called transient jets, are launched. The concurrence of the corona's weakening and the jet's transition from compact to transient implies a connection, but the precise relationship has not been established. Motivated by this, we aim to investigate the corona-jet connection. We perform spectral modeling in the hard and soft X-ray, utilizing \textit{NuSTAR}, \textit{NICER}, and \textit{Swift}/XRT observations to track the evolving X-ray corona for three LMXBs: MAXI\,J1348-630, MAXI\,J1535-571 and MAXI\,J1820+070. We use prior work to mark the presence of compact jets and the dates of discrete jet ejections. We find a clear connection between the evolution of the corona and the jet: across all three sources an increase in the distance of the corona from the black hole occurs near the time that the compact jet is quenched and the transient jet is launched.

\end{abstract}


\section{Introduction} \label{sec:intro}

Black Hole Low-Mass X-ray Binaries (LMXBs) play an important role in understanding the universal phenomenon of accretion/ejection. During outburst cycles, LMXBs evolve rapidly compared to their more massive analogues, supermassive black holes \citep[e.g.,][]{2004cbhg.symp..169M,hopkins2005}, making LMXBs ideal laboratories to study the accretion/ejection phenomena.

As their outburst cycles progresses, LMXBs show characteristic X-ray spectral states, namely the low-hard state (LHS), high-soft state (HSS), and the Intermediate State (IMS). The IMS can be further subdivided into the hard (HIMS) and soft (SIMS) intermediate states based primarily on spectral timing properties \citep[e.g.,][]{Remillard2006, Belloni2010}. LMXB outbursts typically begin in the LHS and transition through the HIMS and SIMS to the HSS before evolving back to the LHS via the IMS. These states are thought to be connected to the geometry of the accretion flow near the black hole \citep[e.g.,][]{Meyer-Hofmeister2009}. During a typical outburst, LMXBs launch two types of jets: compact jets and transient jets, which are distinguished based on their morphology and radio spectrum. During the LHS, partially self-absorbed compact jets produce a flat-to-inverted radio to mm spectrum \citep{Hjellming1988, Harmon1995, Fender2001,Fender2006,gallo_2004,Russell_D_2013,Russell_2014_1836,Tetarenk_2016}, such that the spectral index, $\alpha$, is positive, if $S_\nu \propto \nu^\alpha$, where $S_\nu$ is flux density and $\nu$ is frequency. Near the X-ray peak of the outburst, as the source transitions from HIMS to the SIMS, the compact jet emission is observed to quench \citep{fend04,Russell_2020}, giving way to bright radio flaring, corresponding to the launching of the transient jet \citep{Hjellming95,Fender_1999,Fender2001,car}. The transient jets are composed of discrete knots of synchrotron emitting plasma that are launched out along the jet axis at close to the speed of light \citep{mirabel_2004}. The transient jets exhibit an optically thin radio spectrum, where $\alpha \sim -0.7$ \citep{Fender2006}. Radio flaring is thought to arise as the ejecta interacts with the pre-existing compact jet or the surroundings \citep{Fender2006}.

In the X-ray band, we generally observe back-scattered X-ray emission from the accretion disk, the so-called relativistic X-ray reflection \citep[e.g.,][]{Lightman1988, Ross1993, Bharali2019}. The illuminating source creating the X-ray reflection is commonly referred to as the ``corona", which may be located in the inner part of the disk close to the black hole. The corona is believed to be formed by magnetically heated electrons, with a mean energy of 50 -- 100\,keV \citep[e.g.,][]{haardt91, Haardt1993}. The accretion disk photons gain energy from the corona via inverse Compton scattering \citep{haardt91, Haardt1993}. Despite significant efforts, the location and true nature of the corona remains ambiguous. Constraints have been placed on the coronal size, mainly from X-ray reverberation studies, finding the corona to be compact and located close to the black hole \citep{Kara2016}. Other studies aimed at constraining the geometry of the corona have modeled the RMS-variability and phase-lags of quasi-periodic oscillations (QPOs) observed in LMXB spectra in addition to the time averaged spectra. Using spectral-timing models, \cite{Garcia2021} and \cite{bellavita2022} found that the corona in MAXI\,J1348--630 can be best described as consisting of two components one small and one more extended.  A study of MAXI J1348--630’s QPOs during the rise of the outburst and reflares also suggests a two Component corona, with one being more horizontally extended (along the disk) corona to explain the high fraction emission reflected by the disk \citep{alabarta2025}.
According to \cite{rawat2023}, a two-coronal model is also preferred in the case of MAXI\,J1535--571. Studying the QPOs during the SIMS, \cite{zhang2023} suggests a vertically extended jet-like corona in the SIMS. This more complex and extended coronal geometry, is in agreement with recent results from  (\textit{IXPE}) \cite{ixpe}, which have measured the X-ray polarization of several galactic black holes and thus provided evidence of a corona that is elongated and oriented along the disk \cite[e.g.]{Veledina_2023_ixpe,cyg_ixpe}.

A changing coronal geometry has also been suggested, \cite{ma2023} studied QPOs in MAXI\,J1820+020 in its HIMS-SIMS transition, suggesting that the corona expands horizontally along the disk during the outburst rise and then transforms into a jet-like extended shape in the SIMS,accompanied by a radio flare. \cite{belloni2024} studied QPOs in GRS\,1915+105 to find a rapidly varying corona—with its size varying between $\sim 45$\,$r_g$ and $\sim 270$\,$r_g$ in a single 10.5 hour observation for a $12 M_{\odot}$ BH \cite{reid2023}. 

Early work linking the corona and jet by \cite{gilles1991} suggested a two-phase model in which material ejected from the accretion disk heats the corona and accelerates it to relativistic speeds, producing the jets observed from AGN. Subsequent work on dynamic LMXB systems has continued to link the corona and jet. \cite{Vadawale2003} proposed that the ejection of coronal material may give rise to the transient jets. \cite{markoff_2005} found that the jets could produce some of the emissions attributed to the corona. \cite{koljonen_2015} observed a correlation between the X-ray photon index and properties of the radio jet, suggesting a causal connection between the corona and jet. From X-ray reflection fraction analysis, \cite{you2021} postulated that the corona is a standing shock with material flowing through it. Through analysis of the jet and X-ray spectra of three LMXBs, \cite{cao2022} linked the softening of the X-ray spectrum to changes in the corona-jet system. A study of radio and X-ray data of GRS 1915+105 covering a 10 year period also revealed a correlation between the radio flux from the jet and the strength of the Fe K$\alpha $ line \cite{Mendez2022}, which is thought to arise from the X-ray reflection of coronal emission. 
Several studies performing specto-temporal analysis in the HIMS and SIMS of MAXI\,J1535--571 using \textit{Insight}--HXMT and \textit{NICER} observations have found a vertically extended jet-like corona that contracts before the disappearance of the jet and begins to expand in its absence \citep{Zhang2022, zhang2023, rawat2023}

The process by which particles are ejected via jets in LMXBs is not fully understood, complicating our study of the jet-corona relationship. In the case of compact jets, particles are accelerated into a power-law energy distribution through mechanisms such as shocks or magnetic reconnection \citep{markoff_2001,dalpino_2005,dalpino_2010}. This typically occurs in the jet launching region or at internal shock fronts along the jet \citep{Tetarenko_2019,Fender2004,Malzac2014,Malzac_2012}.
\citet{Vadawale2003} proposed that an ejection of the corona during the transition to the HSS could interact and shock the compact jet and lead to the creation of transient jets. However, the onset of compact jet quenching has been observed during hard-only outbursts, which fail to enter the soft state and a transient jet was not launched \citep[e.g.,][]{Russell_2014,Russell2015}. Nevertheless, it remains possible that the compact jet is quenched by the transient jets, which are created by the ejected corona \citep{Russell_2020}.

In the study presented here, we attempt to better understand the relationship between the corona and the jet during a BH-XRB outburst by comparing their evolution. We perform detailed X-ray spectral modeling to find how the extent of the X-ray corona evolves throughout three LMXB outbursts: MAXI\,J1348--630 (J1348), MAXI\,J1820+070 (J1820), and MAXI\,J1535--571 (J1535). We make use of the radio spectral index ($\alpha$), combined with previously determined transient jet ejection dates, to determine the onset of the transient jet. Section \ref{sec:Obs} outlines and describes the observations used and details our spectral modeling and radio analysis. In section \ref{sec:results}, we briefly outline our results. In section \ref{sec:Disc}.

\section{Observations and Methods}
\label{sec:Obs}
The black hole or black hole candidates J1820, J1348, and J1535 are selected for this study because their recent outbursts are well-observed from radio to X-rays with transient jet ejection dates determined.

We rely on prior work to determine the spectral states and the dates of transition between them. For J1820, we adopt the state transitions dates determined by \cite{Shidatsu_2019}, and identify the HIMS--SIMS transition by the QPO change reported in \citep{Homan2020}. For J1348, we use the state transitions outlined in \cite{zhang}, while for J1535 we take spectral states from \cite{Tao_2018} and \cite{nakahira_2018} compiled by \cite{Russell_2019}.

For our targets in this study, we require both X-ray spectral data and radio observations to track the evolving corona and jet. We focus our efforts on the detailed spectral modeling of the recent X-ray data and rely on published radio data to study the corresponding jet features. As \textit{NuSTAR}'s 3.0-78.0 keV spectral range provides good coverage of the reprocessed spectral features, archival \textit{NuSTAR} observations form the basis of our X-ray spectral analysis, with each \textit{NuSTAR} observation forming an ``\texttt{epoch}". Where possible, we supplement the \textit{NuSTAR} data with \textit{NICER} or \textit{Swift}/XRT data to provide coverage of the soft X-ray energy range and better constrain the disk and reflection features. For epochs with soft X-ray data, we only use the parts of the observations that occur simultaneously. Epoch dates, spectral states, and observations utilized are given in Table \ref{htable}.

\subsection{X-ray Data Reduction}
All X-ray data reduction was performed with Heasoft version 6.32.1, with the CALDB accessed remotely from October through December 2023. 

The \textit{NuSTAR} data reduction was performed with \texttt{NuSTARdas} version 2.2.1. Filtered event files were created using the \texttt{nupipeline} task. The source spectra were extracted from the clean event files after defining a circular region of $\approx85^{\prime\prime}$ centered on the source position. Far from the source location, a similar circular region was used to generate the background spectra. The \texttt{nuproducts} task was used to extract the science products, including auxiliary response files, response matrix files, and energy spectra. If the observations overlapped with \textit{NICER} or \textit{Swift}/XRT data, we ran the \texttt{nuproducts} task with a custom good-time-interval file (GTI). In our analysis, we considered \textit{NuSTAR} spectral data in the energy range $3.0-78.0$\,keV. 

All \textit{NICER} data were reduced using {\tt NICERDAS} version 2023-08-22\_V011a and nicaldbver\_xti\_20221001. Cleaned event files were created from \textit{NICER} data using $nicerl2$. Source spectra and responses were created from the clean event files using the $nicerl3-spect$ command with the specified overlapping GTIs. The background was estimated using the $scorpeon$\footnote{\url{https://heasarc.gsfc.nasa.gov/docs/nicer/analysis_threads/nicerl3-spect/}} background model described in \citet{Remillard2022}. We use the \textit{NICER} spectral data in the $0.8-10.0$\,keV range to avoid several well-studied instrumental and interstellar medium features below this range\footnote{\url{https://heasarc.gsfc.nasa.gov/docs/nicer/data_analysis/workshops/NICER-CalStatus-Markwardt-2021.pdf}}.

For the \textit{Swift}/XRT data analysis, we used \texttt{xrtpipeline} version 0.13.7 to generate clean event files. Further, we employed \texttt{xselect} to trim the event files to times concurrent with \textit{NuSTAR} observations. Pileup was expected for many of the \textit{Swift}/XRT spectra due to the brightness of these sources. To account for this, we used the prescription given by \citet{Romano2006} to remove the brightest source pixels.

Additionally, spectra were also filtered only to include grade 0 events. The \texttt{xrtmkarf} task was then used to create the appropriate response files. Background spectra for the \textit{Swift}/XRT observations were not included in our analysis, as the fields of view of many of these observations did not possess regions free of contamination from the bright sources. We ignore the \textit{Swift}/XRT spectra greater than 4\,keV as \textit{NuSTAR} has higher quality spectral coverage at energies above 4\,keV.

Finally, we used the task \texttt{ftgrouppha} to bin all resulting spectra using a signal-to-noise ratio of 10.

\subsubsection{MAXI\,J1535--571 Excluded X-ray} Data
Initial modeling of our MJD 58010.26 observations of  J1535  led to large residuals in the \textit{NuSTAR} and particularly in \textit{NICER} data. This epoch possesses three overlapping \textit{NuSTAR} and \textit{NICER} orbits. After inspecting the light curve of this observation, we found a sudden increase in the count rates for the two telescopes during the second orbit. While this flare may indeed be real, the study of such flares is not the focus of this paper. Thus, for this epoch, we only used data from the first and third overlapping orbits; this eliminated the residuals. Additionally, large residuals were present for the \textit{NICER} data from MJD 58016.87 to MJD 58034.26 and were not used. Photon pileup could be the reason behind this because pile-up can become an issue in \textit{NICER} above $\sim 2$\,Crab \citep{nicer_pileup}. \texttt{XSPEC} calculations of the {\it NICER} 2--10\,keV flux reveal values significantly above this threshold for these epochs. Various studies reported that J1535 was extraordinarily bright in the soft X-ray band during these epochs \citep[e.g.,][]{Tao_2018}.

\subsection{X-ray Spectral Modeling}

The geometry of the corona--disk is often inferred from the ``reflection" spectrum \citep{Reynolds2003}. Approximating the corona as a compact source located at some height above the black hole (i.e., a lamp-post geometry), changes to this component of the X-ray spectrum can be connected to the corona's distance (or height) from the black hole \citep{dauser}. While the corona is likely more extended, as indicated by recent polarization studies (see discussion), we adopt this simple lamp-post geometry as a probe of the evolution of the Comptonizing region for its ease of use and interpretation. To constrain this height, we model the spectra of each outburst with three components. Firstly, a multi-colour disk blackbody component, described by the \texttt{diskbb} model \citep{diskbb, Makishima1986}. Secondly, we account for the primary X-ray continuum and relativistic reflection from the disk by including the lamp-post \texttt{relxilllp} model version 2.2 \citep{relxill}. Finally, galactic absorption along the line-of-sight is accounted for by the \texttt{tbabs} model \citep{tbabs}. To account for any uncertainty in the cross-instrument calibration, we also include a {\tt constant} model in our modeling that we freeze at one for \textit{NuSTAR}/FPMA. All modeling was performed using \texttt{XSPEC} version 12.13 \citep{xspec} with \texttt{wilm} abundances \citep{wilm} and \texttt{xsect} set to \texttt{vern} \citep{verners}.

Multiple epochs are modeled together within \texttt{XSPEC} to constrain the model parameters better. For J1348, we model all epochs together. Due to computational limitations arising from the large number of epochs in the other two sources, we modeled these epochs in separate groups rather than all together. For J1820, we model our epochs in 3 separate groups, with MJDs 58191.94-58225.30 (initial LHS) constituting group 1, MJDs 58242.27-58327.05 (covering the end of the LHS, the HIMS and SIMS, and the beginning of the HSS) comprising group 2, and MJDs 58343.61-58404.95 (final LHS) making group 3.  
For J1535, we split our epochs into two groups. Here, group 1 covers the initial LHS and most of the transition to the HSS and consists of MJDs 58004.35-58017.21. The remaining observations (MJDS 58023.11-58090.77) cover the transition back to the LHS and then to the HSS and form group 2.

We consider the black hole spin, the iron abundance, the inclination angle of the disk, and the column density to be ``global" parameters, which we leave free but tie together between all epochs in each group. These global parameters were allowed to vary between groups. We discuss the impact of this further in section \ref{sec:results}.

In the \texttt{relxilllp} model, we let the model return the primary continuum and the expected reflection spectrum for a lamp post geometry, while \texttt{diskbb} accounts for the temperature and normalization of the accretion disk.  The disk temperature, disk normalization, corona height ($h$), inner-disc radius, the power law index, the disk ionization, the cutoff energy, and \texttt{relxilllp} normalization were all left free to vary between epochs but were linked between observations within a given epoch.

We estimated the errors of the best-fit parameters using the Markov Chain Monte Carlo (MCMC) simulations. For each group of epochs, we use \texttt{XSPEC}'s built-in Goodman-Weare MCMC algorithm with 270 walkers and a chain length of 6 million elements with the first 2 million elements burned. Errors for each parameter were estimated from the resulting chains by sorting the chain values and taking the central 90\%. The numerical results of our X-ray spectral modeling can be found in Table~\ref{htable}.

\subsubsection{MAXI\,J1820+070 Additional Blackbody Component}
Our initial model failed to reproduce the observed spectra in several J1820 epochs. Excess residuals between our data and this model were observed in both \textit{NuSTAR} and \textit{NICER} spectra below $10.0$\,keV our observations on MJDs 58201.53, 58201.86, 58224.96, 58259.33, 58297.20, 58315.02, 58349.82, and 58389.35. Following the analysis of the same data by \cite{Fabian_2020}, we add a blackbody component to our model for this source. This extra blackbody component likely arises from disk material emitting from or within the innermost stable circular orbit (ISCO), e.g. the plunging region \citep{1820_isco, Fabian_2020}. While a more robust model accounting for material within the plunging region has been developed by \cite{1820_isco}, a complete analysis with such a model is beyond the scope of this paper. 

\subsection{Analysis of Jet Features}
We use the radio fluxes from prior work for all three sources to track the jet behavior. For J1348 and J1535, we also use the radio flux densities and spectral indices, $\alpha$, presented by \citet{car} and \citet{Russell_2019}, respectively. For J1820, radio spectral indices were calculated by fitting a power law to the quasi-simultaneous radio flux values (taken from a variety of radio telescopes in the 0.1365-350 GHz range) reported by \citep{atel11887,Bright_2020,Echibur_Trujillo_2024,Shaw_2021,atri2020} and compiled in \cite{Echibur_Trujillo_2024}. Throughout this work, we interpret a shift from optically thick to thin values of $\alpha$ as a quenching of the compact jet, as it would be expected as part of a transition to the transient jet. We note that although the compact jet is quenched during the HSS, steep-spectrum radio emission can linger at the core of the outburst deep into the HSS \citep[e.g.,][]{Russell_2019, hughes_2025a}. For J1535, this core radio emission faded $\sim$\,5 months into the HSS, while in the case of the black hole low-mass X-ray binary Swift\,J1727.8--1613, the radio emission from the core remained until the transition back to the LHS \citep{hughes2025b}. This lingering emission is often attributed to transient ejecta that have not traveled far enough to be resolved from the core position.

To determine the dates of the launching of discrete jet ejecta we rely on values reported in the literature. Jet launch dates for J1535, and J1820 are taken from \citet{Russell_2019} and \citet{Wood_2021} respectively. In the case of MAXI\,J1348-630, there were multiple excursions to the IMS during the HSS \citep{car,zhang}. One of these excursions launched a second ejection event that became resolved from the compact source. \cite{car} detected both the first and second ejection events as they propagated away from the compact object.


\section{Results}
\label{sec:results}
\begin{figure}[ht!]
\centering
\includegraphics[width=\columnwidth]{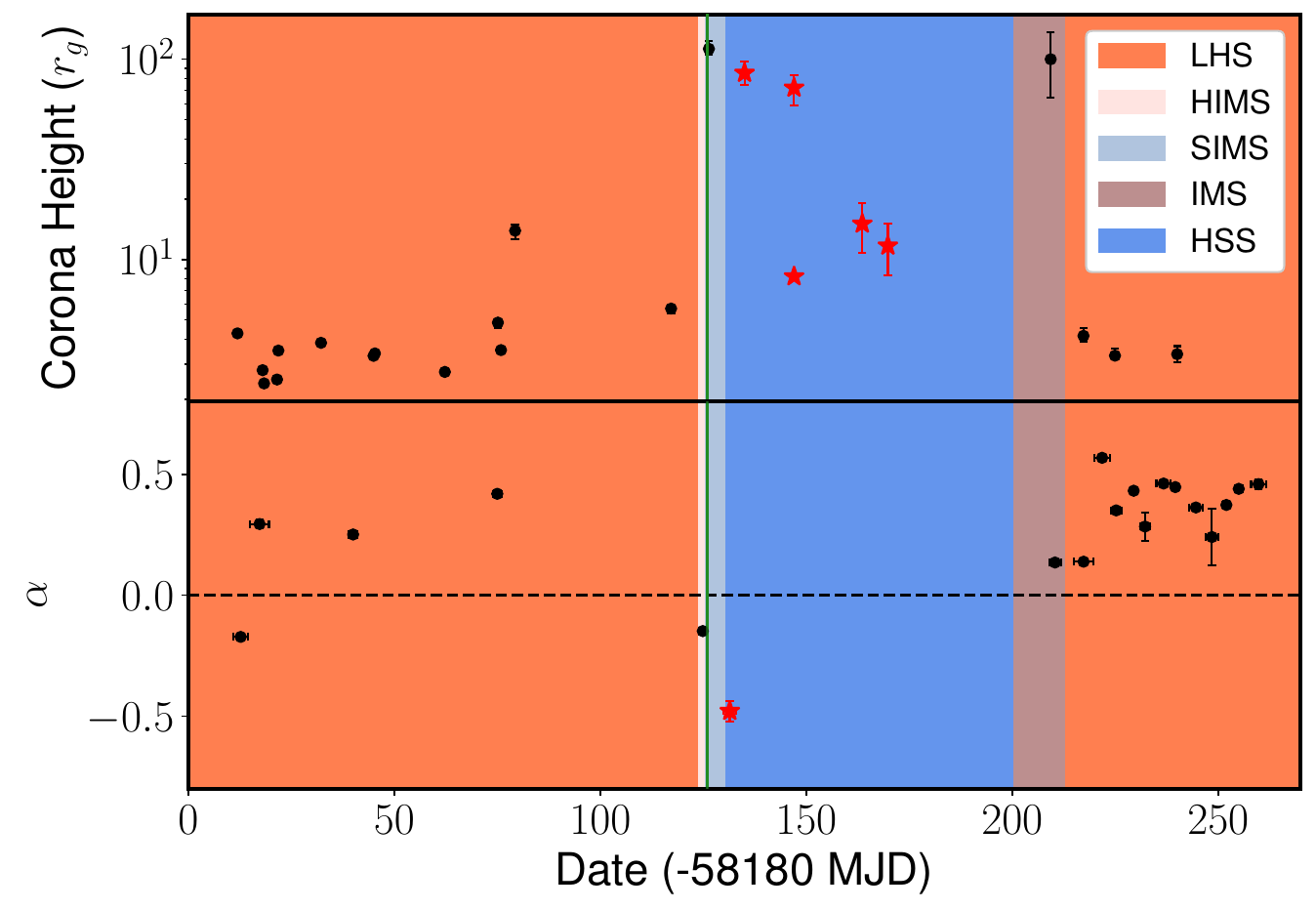}
\caption{Top Panel: The corona heights of MAXI J1820+070's 2018 outburst determined through X-ray spectral modeling. Bottom Panel: The evolution of the jet's radio spectral index during the same outburst. For this source we determine the radio spectral index by modeling radio flux values compiled by \citep{Echibur_Trujillo_2024} with a simple power law. Modeled heights and values of $\alpha$ change from black to red when $\alpha$ dips below $-0.25$ indicating a move towards optically thin values for the jet. Background shading represents the source's spectral states taken from \cite{Shidatsu_2019} and \cite{Homan2020}. The vertical green line present in both panels indicates the date of the jet ejection according to \cite{Wood_2021}. Errors on heights are estimated from our MCMCs, and are reported at 90\% confidence for all epochs. In this and following figures, a radio measurement with no reported corona size indicates a lack of available \textit{NuSTAR} data on that date. 
\label{m1820_h}}
\end{figure}

\begin{figure}[ht!]
\centering
\includegraphics[width=\columnwidth]{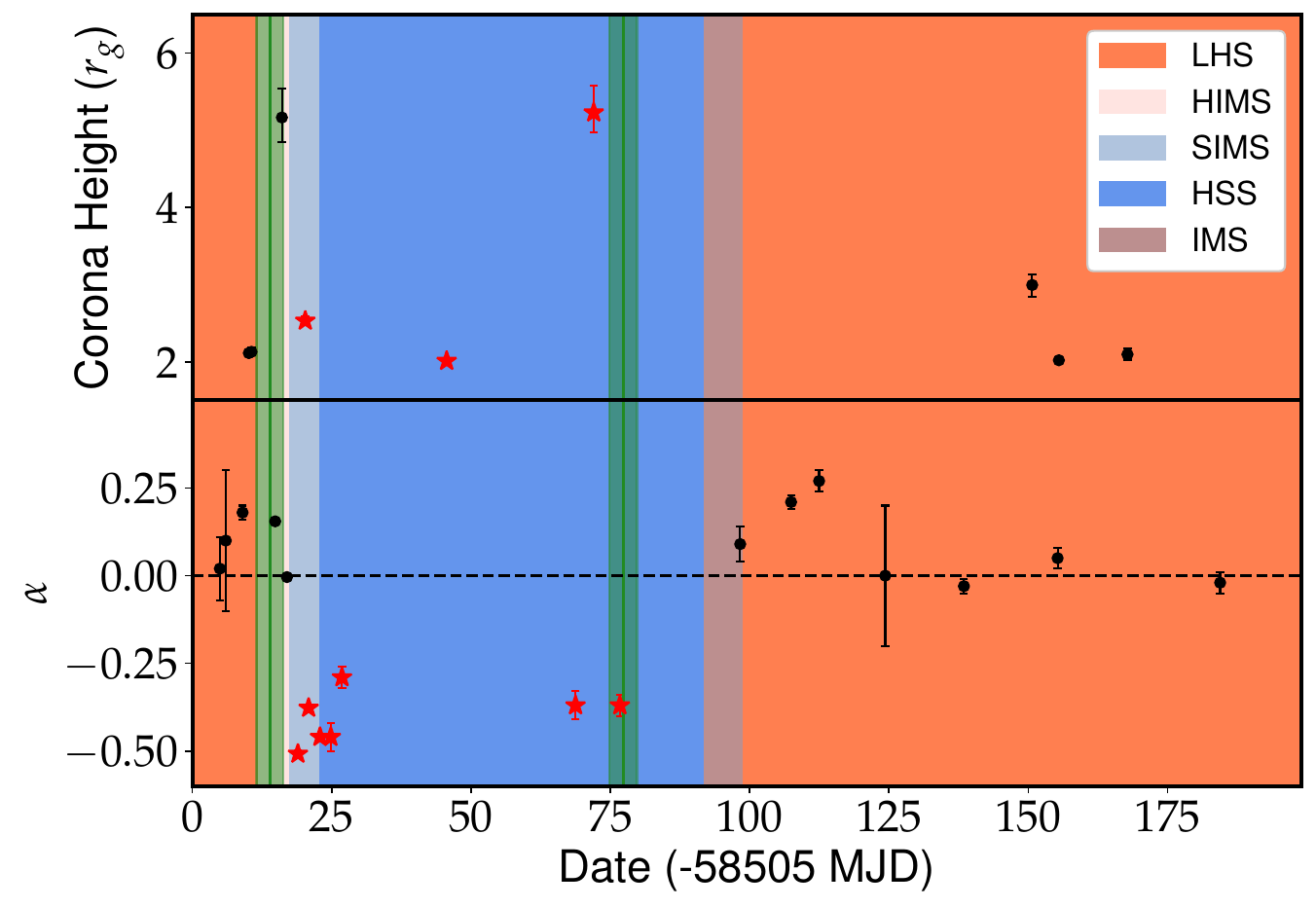}
\caption{Top Panel: Observed corona heights of MAXI\,J1348--630 during its 2019 outburst, with background shading indicating the dates of spectral state transitions according to \cite{zhang}. Bottom Panel: The radio spectral indices obtained by  \cite{car}. Values of $\alpha$ and corona height are shaded red when $\alpha < -0.25$. An increased corona height is observed around the time of both ejection events (indicated by green shading) \citep{car}.
\label{1348_h}}
\end{figure}

\begin{figure}[ht!]
\centering
\includegraphics[width=\columnwidth]{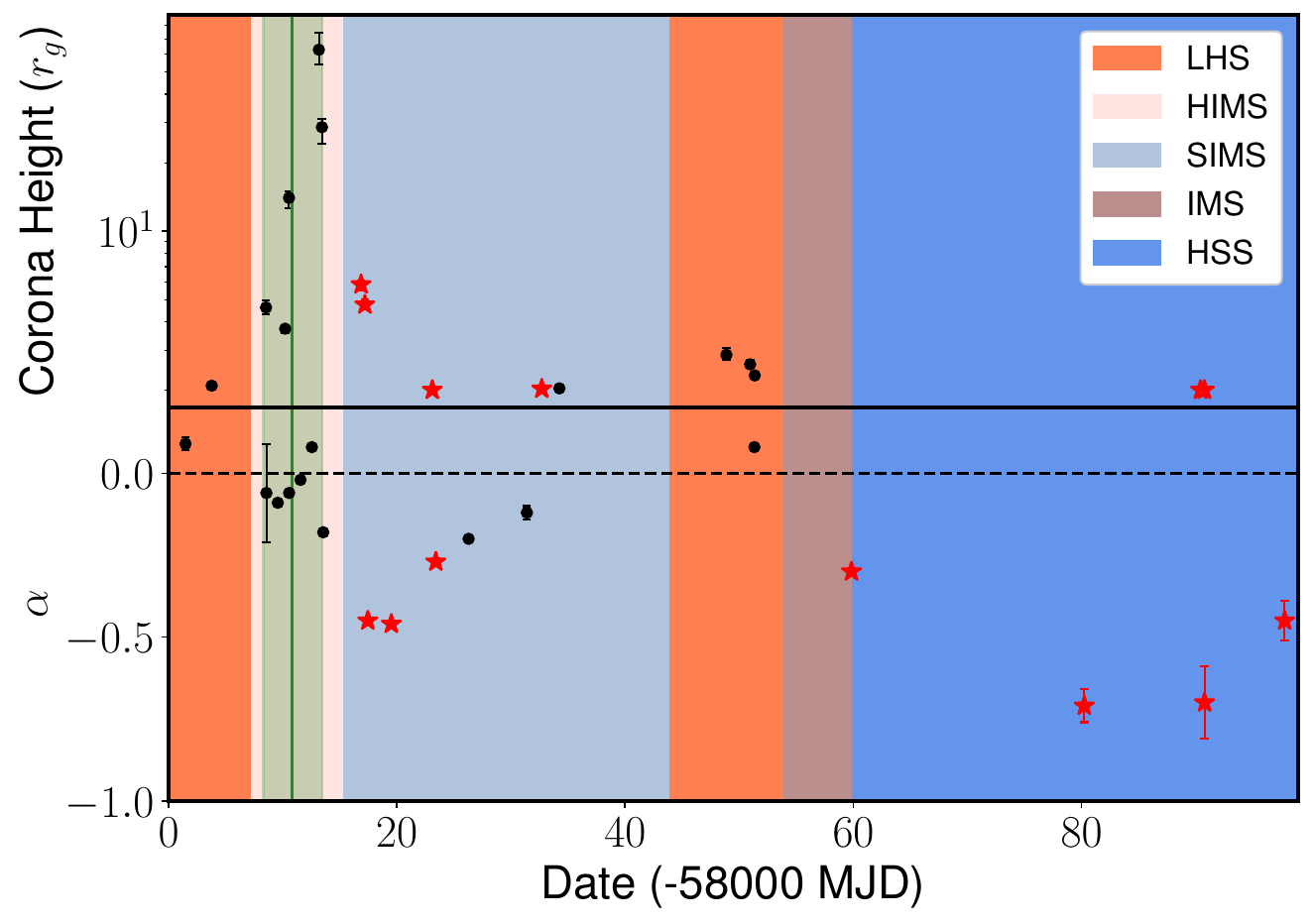}
\caption{Top Panel: Observed corona heights of MAXI J1535\--571 during its 2017 outburst. Bottom panel: The radio/IR spectral indices from \cite{Russell_2019} for the same time period. An increase in corona at the time of jet ejections and the quenching of the compact jet is observed. Values of $\alpha$ and corona height are shaded red once the radio spectrum is $\alpha < -0.25$. Background shading represents spectral states taken from \cite{Tao_2018,nakahira_2018,Russell_2019}. Green shading represents dates of jet ejection \citep{Russell_2019}. 
\label{m1535_h}}
\end{figure}

Our X-ray spectral modeling aimed to constrain the height of the corona above the accretion disk and examine its correlation with the jet's behavior. Figures \ref{m1820_h}, \ref{1348_h}, and \ref{m1535_h} show the variation of the coronal height and $\alpha$ with time. In all three sources, the corona's height maintains a low, nearly constant value during the LHS (see Table~\ref{htable}). However, around the time when the compact jet is quenched, and a transient jet is launched, the corona height increases markedly.

The results for J1820 are visualized in Figure \ref{m1820_h}. During its long initial LHS, the corona height hovers at relatively low values, between $2.4$ and $4.8$\,$r_g$ until MJD 58259 when it jumps up to $14\pm 1$\,$r_g$ before returning to $5.7 \pm 0.3$\,$r_g$ by MJD 58297. Following this, the inferred height of the corona increases to $113^{+11}_{-7}$\,$r_g$ on MJD 58306. This happens concurrently with the radio spectrum becoming more optically thin as $\alpha$ decreases to $-0.15 \pm 0.01$ on MJD 58305, then to $-0.48\pm0.04$ on MJD 58311. These changes also coincide with the jet ejection event on MJD 58306 \citep{Wood_2021}. The corona height remains high for the following two epochs before moving back toward the black hole as the HSS progresses. During the transition to the second LHS, the compact jet is detected again, and corona height reaches $\sim 100$\,$r_g$. In the final LHS, as $\alpha$ recovers, the corona height returns to low values of about $4$\,$r_g$. 
\vspace{1.0em}

A similar pattern is observed in J1348 (Figure \ref{1348_h}). Our observations begin in the initial LHS with the corona height low ($\sim 2$\,$r_g$). The corona height subsequently increases to $5.2 \pm 0.4$\,$r_g$. on MJD 58521.07 The increase in coronal height again coincides with $\alpha$'s decrease from low positive to steep negative values and the ejection of the jet \citep{car}. The following observations see the corona return to lower values. In this source, the corona height is modeled at $5.2 \pm 0.3$\,$r_g$ during MJD 58577.03; this immediately precedes the projected date of a second ejection of plasma blob \citep{car}. As $\alpha$ returns to positive values during the final LHS, we measure low coronal heights again.

J1535 (Figure \ref{m1535_h}) also exhibits a low corona height during the initial LHS. The height quickly increases to a maximum height of $63^{+13}_{-9}$\,$r_g$ as $\alpha$ decreases to negative values \citep{Russell_2019}. This increase in coronal height again occurs around the time of the jet ejection \citep{chauhan_2021,Russell_2019}. The corona height falls to low values again for the remainder of our epochs. 

In J1820 and J1535, the decision to separate our epochs into groups led to a different set of ``global" parameters for each group. There are notable variations in some of these parameters between different spectral groupings of the same source. Most notable of these differences is the gradual change in inclination through the three groups of J1820 from $\sim$\,26$\degree$ in group 1 to $\sim$\,30$\degree$ in group 2 to $\sim$\,47$\degree$ in group 3. Fitting was initially performed with the inclination and other global parameters of groups 1 and 3 frozen to best fit values of the those in group 2. This resulted in a poorer $\chi^2$, but still preserved our reported trend in coronal height changes. While significant changes to the inclination of LMXB GRS\,1915+105 have been observed to occur within a year \citep{Rodriguez_2025}, this is not something generally observed or expected in LMXBs. Another global parameter exhibiting changes between groups is the \texttt{tbabs} equivalent hydrogen column density parameter nH, which goes from about 0.4\,$\times10^{22}$\,atoms\,cm$^{-2}$ in group 1 to about 0.07$\times10^{22}$\,atoms\,cm$^{-2}$ in groups 2 and 3 of J1820. A version of our J1820 group 1 fits with the nH parameter fixed to 0.07$\times10^{22}$ was performed with results shown in the appendix alongside our reported results. This resulted in a poorer $\chi^2$, with no significant change to the corona heights. We note that the number of epochs using supplementary \textit{NICER} or \textit{Swift}/XRT data differs between groups, potentially contributing to differences in some of the global parameters. However, this is unlikely to affect the observed trend in corona heights, as illustrated by our attempts to mitigate the global parameter differences. Additionally, the major increases and decreases in corona height are seen to occur within our individual spectral groups.

\vspace{1.0cm}

\section{Discussion}
\label{sec:Disc}
Our results suggest a relationship between the jet and the corona, revealing a change in the corona geometry contemporaneous with a jet ejection event. As outlined in section \ref{sec:results}, the main outcomes of our work are the following:

1) In the initial LHSs, we observed that the X-ray corona remains steadily close to the black hole, and the compact radio jet is also observed to remain mostly unchanged.

2) As the LMXB transitions from the LHS to the HSS, we find that the corona height increases dramatically. Concurrently, the compact jet emission is quenched, and transient jet ejecta are launched from the system. This change in jet behavior is identified through spatially resolved ejecta and the emergence of optically thin synchrotron emission, characteristic of the short-lived transient jet before the jet disappears entirely as the source enters the HSS.

3) In J1820, as the outburst fades and the source returns to a hard state, the compact jet recovers \citep{Echibur_Trujillo_2024}, the corona height once again increases before returning to the its original low levels at later times.

The idea of a connection between the radio jet and the corona is not new; for over two decades it has been postulated that the jet or jet base is connected to the corona and responsible for the relativistic reflection features in X-ray binaries \citep{Markoff2001, Markoff2003, Markoff2004}. In a model proposed by \citet{Markoff2001}, it has been argued that inverse Compton and synchrotron emission from the base of the jet could be contributing to the non-thermal power-law spectra or even produce the detected X-ray reflection \citep{Markoff2004}. Alternatively, it has been proposed that the Compton-scattering of soft photons could occur entirely within the jet during the LHS \citep{Reig2003,Giannios2004,kylafis2008}, and that the lingering power-law component seen during the HSS could be produced by a cool disk extending all the way to the ISCO rather than a corona \citep{Reig2001}. Comptonization by the jet rather than a corona has been somewhat successful in explaining the reflection component \citep{Reig2021}, however this model predicts a flatter emissivity than what reflection model fits typically find \citep{uttley2025}. \cite{Wang_2020hh} suggested that there are two Comptonization regions, with one being the more compact corona near the black hole suggested by \cite{kara19}, and the other occurring within the jet. 

Various X-ray spectral and timing studies have already noted that the corona is dynamic in nature, i.e., the corona geometry changes with the accretion rate change \citep[e.g.,][]{kara19, Garc_a_2020, Lucchini2021,alabarta2025,rawat2023,ma2023}. Studies of QPOs in LMXBs have revealed much about the dynamics of the corona. Fast variations in the QPO time lags of GRS\,1915+105 suggest that the size of the corona can vary by nearly an order of magnitude over the course of a single 10.5 hour observation \citep{belloni2024}.  However, the exact nature of the corona evolution remains unclear. \cite{kara19} studied reverberation lags in J1820 during its initial LHS and found that the corona gradually contracted as the spectrum softened. We find that our results do not align with those reported by \cite{kara19}. This discrepancy may arise because the two studies are probing different regions of the corona. While \cite{kara19} observed significant variability of the lags, which they interpreted as reverberation lags, the iron K emission line profile—on which our modeling is based—remains relatively stable within \textit{NICER}’s energy band.

\cite{Lucchini2021} provided further evidence for a dynamic corona and suggested a link to the jet-base. In MAXI\,J1836--194, they found that harder states ($\Gamma \lesssim 2.0$) require a more compact corona/jet-base region close to the black hole, while softer states require acceleration to move downstream. Further studies of quasi-periodic oscillations evolution led some to suggest a dual corona model \citep{Garcia2021}, that is sometimes also seen as a vertically extended corona/jet ecosystem that varies in size. \citep{zhang2023, sripada2022, liao2024, Liu_2022, Garcia2022, rawat2023, Mendez2022, ma2023}. For example, \cite{zhang2023} found a vertically extended corona varying from 100\,$r_g$ to 700\,$r_g$ during the HIMS-SIMS transition of J1535 (assuming a $10 M_{\odot}$ BH). \cite{liao2024} found corona sizes that varied between $\sim 200$\,$r_g$ and $\sim 700$\,$r_g$ in Swift\,J1727.8--1613 (for a BH of size $\sim M_{\odot}$ \citep{sanchez2025}). In a \textit{NICER} spectral timing study of J1820 \cite{wang2021} found increasing soft reverberation lags during the hard-to-soft transition, a change they attribute to the corona expanding from $\sim 42$\,$r_g$ to $>335$\,$r_g$. This trend of an expanding corona is consistent with our results from the same time period, and they suggest this expansion could result in the launching of a jet knot.  Alternatively, these increasing lags could also be understood in the context of an evolving accretion disk atmosphere, although the thermalization timescale is often very small \citep{salveson2022} 

To account for the effects of an extended corona on the observed spectrum, \cite{Lucchini2023} utilized a dual corona model, originally proposed by \cite{Garcia2021}, in which a lower component (several $r_g$) dominates the features observed in the time-average spectrum, while the second component, a higher corona (hundreds of $r_g$) primarily accounts for the time-variable features. \citet{Lucchini2023} tested this model on J1820 ({\it NICER} observation on MJD 58301). The authors found a lower corona at height $\sim 20$\,$r_g$, and a much fainter upper corona at height $\sim 280$\,$r_g$. That {\it NICER} observation occurs between our observations on MJDs 58297 and 58306, where our single-lamppost corona height went from $\sim 5$\,$r_g$ to $\sim 65$\,$r_g$. Our simpler model is consistent with these results if our single lamp-post corona height can be seen as an flux-weighted average position of the two coronal positions. The success of their model in explaining both the time-variability and the time-averaged X-ray spectrum points to a likely scenario: that while a core of the corona resides somewhat close to the black hole, elements of it stretch out to greater distances. Both their model and ours act as probes of this more extended corona.

During the LHS, the corona height always remained at a low value or nearly constant (see Table~\ref{htable}), and the radio spectrum is consistently flat or inverted. This suggests that the geometry of the corona is close to the black hole at this time. \cite{Russell_2020} found that the compact particle-acceleration region of J1535 lies at distances of thousands of $r_g$ from the black hole during the LHS. \cite{Echibur_Trujillo_2024} found that this region lies even further away during J1820's LHS. This indicates that the corona and the jet particle acceleration region are two separate entities.

However, we observed an increase in the corona height around the time of the transient jet ejection, suggesting a connection between the two. We note that the transient jet is launched around the same time as the compact jet quenching in our three sources; as such, due to the low cadence of our X-ray data, we cannot precisely identify what the immediate connection is. We do see an increase in the coronal height at the same time as the compact jet is re-established in J1820, suggesting that the connection may be with the compact jet. Either way, this work implies that the corona and the outflow are connected through some physical process. Considering our findings and following the models of \citet{Markoff2001, Markoff2004}, we argue that the corona and the jet acceleration region are coupled, with continuous feedback between them \citep{Malzac2014}. A magnetic field likely originating from the accretion disk, or threaded in the event horizon could be the mechanism behind this coupling. Parts of the corona could be ejected out far beyond the lamp-post height assumptions and possibly interact with the first acceleration region or jet base through the magnetic field lines \citep{Narayan2003, Igumenshchev2008, Tchekhovskoy2011, Janiuk2022}.

An alternative explanation could be that a change in the accretion rate close to the LHS to HSS  transition (point of transient jet eject) makes fast-moving matter clump into slow-moving plasma, creating a shock in the accretion disk \citep{Fender2004a, Narayan2012}. This shock may change the accompanying magnetic field configuration and lead to a change in the corona geometry, possibly from compact to the lamp-post structure, which leads to a change in the coronal height.

This also suggests that the corona might have a more complex geometry, and it could be a combination of both an extended corona and the near-point source. Rapid changes to the coronal geometry as the jet becomes transient have already been observed in Swift\,J1727.8--1613 \citep{Podgorny2024}. The authors also reported a return to pre-outburst polarization levels upon re-entering the LHS, indicating that the coronal geometry and nature were similar to those before the outburst \citep{Podgorny2024}. Furthermore, a drastic decrease in polarization during the HSS has been observed \citep{Mastroserio2024, Soboda2024}. The polarization findings also lend support to the notion of a non-trivial coronal geometry; a radially extended portion is especially indicated by polarization \citep{Veledina_2023_ixpe, cyg_ixpe, Ingram_2024, Steiner2024} However, it should be noted that interpretation of the polarization measurements hinges on simulation-based models that only consider certain geometries. Although some reverberation studies study suggest a small corona, others have also suggested the possibility of a corona that is radially extended along the disk \citep{ma2023,alabarta2025}. Reconciling these findings with spectral modeling results suggesting a more compact or vertically extended coronal geometry is challenging. It has been suggested \citep[e.g.][]{alabarta2025} that the corona both begins and ends an outburst with a compact component and a large, horizontally extended component, while the large component takes on a jet-like shape at the peak of the outburst.

Our analysis cannot determine whether the jet ejection or the coronal height increase occurred first or whether the corona is entwined with the compact jet or the transient jet. Jet ejection times are difficult to pinpoint \citep{Wood_2021, Russell_2020,car}. Additionally, our X-ray spectral epochs are too irregular to track the exact onset of coronal height changes, making it challenging to identify the cause of these geometric changes. Higher-cadence simultaneous observations in radio and X-ray bands could clarify whether the coronal height increase precedes jet ejection, providing new insights.

\section*{Acknowledgements}
We sincerely thank the referee for their thorough and constructive review. We appreciate the time and effort they put into evaluating our manuscript, and we have carefully considered all comments and suggestions.
This work is partly supported by the NSF REU Program (Award Number 2244344) at Montana State University. 
This work is additionally supported by the Montana Space Grant Consortium (Grant Number 80NSSC20M0042).
JC and RS acknowledge support from the Leverhulme Trust grant RPG-2023-240.

\bibliography{main_paper}{}

@ARTICLE{Lucchini2021,
       author = {{Lucchini}, M. and {Russell}, T.~D. and {Markoff}, S.~B. and {Vincentelli}, F. and {Gardenier}, D. and {Ceccobello}, C. and {Uttley}, P.},
        title = "{Correlating spectral and timing properties in the evolving jet of the microblazar MAXI J1836-194}",
      journal = {\mnras},
         year = 2021,
        month = mar,
       volume = {501},
       number = {4},
        pages = {5910-5926},
          doi = {10.1093/mnras/staa3957},
archivePrefix = {arXiv},
       eprint = {2012.14967},
 primaryClass = {astro-ph.HE},
       adsurl = {https://ui.adsabs.harvard.edu/abs/2021MNRAS.501.5910L}
}

@ARTICLE{Lucchini2023,
       author = {{Lucchini}, Matteo and {Mastroserio}, Guglielmo and {Wang}, Jingyi and {Kara}, Erin and {Ingram}, Adam and {Garcia}, Javier and {Dauser}, Thomas and {van der Klis}, Michiel and {K{\"o}nig}, Ole and {Lewin}, Collin and {Nathan}, Edward and {Panagiotou}, Christos},
        title = "{Investigating the Impact of Vertically Extended Coronae on X-Ray Reverberation Mapping}",
      journal = {\apj},
         year = 2023,
        month = jul,
       volume = {951},
       number = {1},
          eid = {19},
        pages = {19},
          doi = {10.3847/1538-4357/acd24f},
archivePrefix = {arXiv},
       eprint = {2305.05039},
 primaryClass = {astro-ph.HE},
       adsurl = {https://ui.adsabs.harvard.edu/abs/2023ApJ...951...19L}
}

@ARTICLE{Steiner2024,
       author = {{Steiner}, James F. and {Nathan}, Edward and {Hu}, Kun and {Krawczynski}, Henric and {Dov{\v{c}}iak}, Michal and {Veledina}, Alexandra and {Muleri}, Fabio and {Svoboda}, Jiri and {Alabarta}, Kevin and {Parra}, Maxime and {Bhargava}, Yash and {Matt}, Giorgio and {Poutanen}, Juri and {Petrucci}, Pierre-Olivier and {Tennant}, Allyn F. and {Baglio}, M. Cristina and {Baldini}, Luca and {Barnier}, Samuel and {Bhattacharyya}, Sudip and {Bianchi}, Stefano and {Brigitte}, Maimouna and {Cabezas}, Mauricio and {Cangemi}, Floriane and {Capitanio}, Fiamma and {Casey}, Jacob and {Rodriguez Cavero}, Nicole and {Castellano}, Simone and {Cavazzuti}, Elisabetta and {Chun}, Sohee and {Churazov}, Eugene and {Costa}, Enrico and {Di Lalla}, Niccol{\`o} and {Di Marco}, Alessandro and {Egron}, Elise and {Ewing}, Melissa and {Fabiani}, Sergio and {Garc{\'\i}a}, Javier A. and {Green}, David A. and {Grinberg}, Victoria and {Hadrava}, Petr and {Ingram}, Adam and {Kaaret}, Philip and {Kislat}, Fabian and {Kitaguchi}, Takao and {Kravtsov}, Vadim and {Kub{\'a}tov{\'a}}, Brankica and {La Monaca}, Fabio and {Latronico}, Luca and {Loktev}, Vladislav and {Malacaria}, Christian and {Marin}, Fr{\'e}d{\'e}ric and {Marinucci}, Andrea and {Maryeva}, Olga and {Mastroserio}, Guglielmo and {Mizuno}, Tsunefumi and {Negro}, Michela and {Omodei}, Nicola and {Podgorn{\'y}}, Jakub and {Rankin}, John and {Ratheesh}, Ajay and {Rhodes}, Lauren and {Russell}, David M. and {{\v{S}}lechta}, Miroslav and {Soffitta}, Paolo and {Spooner}, Sean and {Suleimanov}, Valery and {Tombesi}, Francesco and {Trushkin}, Sergei A. and {Weisskopf}, Martin C. and {Zane}, Silvia and {Zdziarski}, Andrzej A. and {Zhang}, Sixuan and {Zhang}, Wenda and {Zhou}, Menglei and {Agudo}, Iv{\'a}n and {Antonelli}, Lucio A. and {Bachetti}, Matteo and {Baumgartner}, Wayne H. and {Bellazzini}, Ronaldo and {Bongiorno}, Stephen D. and {Bonino}, Raffaella and {Brez}, Alessandro and {Bucciantini}, Niccol{\`o} and {Chen}, Chien-Ting and {Ciprini}, Stefano and {De Rosa}, Alessandra and {Del Monte}, Ettore and {Di Gesu}, Laura and {Donnarumma}, Immacolata and {Doroshenko}, Victor and {Ehlert}, Steven R. and {Enoto}, Teruaki and {Evangelista}, Yuri and {Ferrazzoli}, Riccardo and {Gunji}, Shuichi and {Hayashida}, Kiyoshi and {Heyl}, Jeremy and {Iwakiri}, Wataru and {Jorstad}, Svetlana G. and {Karas}, Vladimir and {Kolodziejczak}, Jeffery J. and {Liodakis}, Ioannis and {Maldera}, Simone and {Manfreda}, Alberto and {Marscher}, Alan P. and {Marshall}, Herman L. and {Massaro}, Francesco and {Mitsuishi}, Ikuyuki and {Ng}, Chi-Yung and {O'Dell}, Stephen L. and {Oppedisano}, Chiara and {Papitto}, Alessandro and {Pavlov}, George G. and {Peirson}, Abel L. and {Perri}, Matteo and {Pesce-Rollins}, Melissa and {Pilia}, Maura and {Possenti}, Andrea and {Puccetti}, Simonetta and {Ramsey}, Brian D. and {Roberts}, Oliver J. and {Romani}, Roger W. and {Sgr{\`o}}, Carmelo and {Slane}, Patrick and {Spandre}, Gloria and {Swartz}, Douglas A. and {Tamagawa}, Toru and {Tavecchio}, Fabrizio and {Taverna}, Roberto and {Tawara}, Yuzuru and {Thomas}, Nicholas E. and {Trois}, Alessio and {Tsygankov}, Sergey S. and {Turolla}, Roberto and {Vink}, Jacco and {Wu}, Kinwah and {Xie}, Fei},
        title = "{An IXPE-led X-Ray Spectropolarimetric Campaign on the Soft State of Cygnus X-1: X-Ray Polarimetric Evidence for Strong Gravitational Lensing}",
      journal = {\apjl},
         year = 2024,
        month = jul,
       volume = {969},
       number = {2},
          eid = {L30},
        pages = {L30},
          doi = {10.3847/2041-8213/ad58e4},
archivePrefix = {arXiv},
       eprint = {2406.12014},
 primaryClass = {astro-ph.HE},
       adsurl = {https://ui.adsabs.harvard.edu/abs/2024ApJ...969L..30S}
}

@ARTICLE{Mastroserio2024,
       author = {{Mastroserio}, G. and {De Marco}, B. and {Baglio}, M.~C. and {Carotenuto}, F. and {Fabiani}, S. and {Russell}, T.~D. and {Capitanio}, F. and {Cavecchi}, Y. and {Motta}, S. and {Russell}, D.~M. and {Dovciak}, M. and {Del Santo}, M. and {Alabarta}, K. and {Ambrifi}, A. and {Campana}, S. and {Casella}, P. and {Covino}, S. and {Illiano}, G. and {Kara}, E. and {Lai}, E.~V. and {Lodato}, G. and {Manca}, A. and {Mariani}, I. and {Marino}, A. and {Miceli}, C. and {Saikia}, P. and {Shaw}, A.~W. and {Svoboda}, J. and {Vincentelli}, F.~M. and {Wang}, J.},
        title = "{X-ray and optical polarization aligned with the radio jet ejecta in GX 339-4}",
      journal = {arXiv e-prints},
         year = 2024,
        month = aug,
          eid = {arXiv:2408.06856},
        pages = {arXiv:2408.06856},
          doi = {10.48550/arXiv.2408.06856},
archivePrefix = {arXiv},
       eprint = {2408.06856},
 primaryClass = {astro-ph.HE},
       adsurl = {https://ui.adsabs.harvard.edu/abs/2024arXiv240806856M}
}

@ARTICLE{Soboda2024,
       author = {{Svoboda}, Ji{\v{r}}{\'\i} and {Dov{\v{c}}iak}, Michal and {Steiner}, James F. and {Kaaret}, Philip and {Podgorn{\'y}}, Jakub and {Poutanen}, Juri and {Veledina}, Alexandra and {Muleri}, Fabio and {Taverna}, Roberto and {Krawczynski}, Henric and {Brigitte}, Ma{\"\i}mouna and {Datta}, Sudeb Ranjan and {Bianchi}, Stefano and {Mu{\~n}oz-Darias}, Teo and {Negro}, Michela and {Rodriguez Cavero}, Nicole and {Castro Segura}, Noel and {Bollemeijer}, Niek and {Garc{\'\i}a}, Javier A. and {Ingram}, Adam and {Matt}, Giorgio and {Nathan}, Edward and {Weisskopf}, Martin C. and {Altamirano}, Diego and {Baldini}, Luca and {Capitanio}, Fiamma and {Egron}, Elise and {Emami}, Razieh and {Hu}, Kun and {Marra}, Lorenzo and {Mastroserio}, Guglielmo and {Petrucci}, Pierre-Olivier and {Ratheesh}, Ajay and {Soffitta}, Paolo and {Tombesi}, Francesco and {Yang}, Yi-Jung and {Zhang}, Yuexin},
        title = "{Dramatic Drop in the X-Ray Polarization of Swift J1727.8{\textendash}1613 in the Soft Spectral State}",
      journal = {\apjl},
         year = 2024,
        month = may,
       volume = {966},
       number = {2},
          eid = {L35},
        pages = {L35},
          doi = {10.3847/2041-8213/ad402e},
archivePrefix = {arXiv},
       eprint = {2403.04689},
 primaryClass = {astro-ph.HE},
       adsurl = {https://ui.adsabs.harvard.edu/abs/2024ApJ...966L..35S}
}

@ARTICLE{Podgorny2024,
       author = {{Podgorn{\'y}}, J. and {Svoboda}, J. and {Dov{\v{c}}iak}, M. and {Veledina}, A. and {Poutanen}, J. and {Kaaret}, P. and {Bianchi}, S. and {Ingram}, A. and {Capitanio}, F. and {Datta}, S.~R. and {Egron}, E. and {Krawczynski}, H. and {Matt}, G. and {Muleri}, F. and {Petrucci}, P. -O. and {Russell}, T.~D. and {Steiner}, J.~F. and {Bollemeijer}, N. and {Brigitte}, M. and {Castro Segura}, N. and {Emami}, R. and {Garc{\'\i}a}, J.~A. and {Hu}, K. and {Iacolina}, M.~N. and {Kravtsov}, V. and {Marra}, L. and {Mastroserio}, G. and {Mu{\~n}oz-Darias}, T. and {Nathan}, E. and {Negro}, M. and {Ratheesh}, A. and {Rodriguez Cavero}, N. and {Taverna}, R. and {Tombesi}, F. and {Yang}, Y.~J. and {Zhang}, W. and {Zhang}, Y.},
        title = "{Recovery of the X-ray polarisation of Swift J1727.8{\ensuremath{-}}1613 after the soft-to-hard spectral transition}",
      journal = {\aap},
         year = 2024,
        month = jun,
       volume = {686},
          eid = {L12},
        pages = {L12},
          doi = {10.1051/0004-6361/202450566},
archivePrefix = {arXiv},
       eprint = {2404.19601},
 primaryClass = {astro-ph.HE},
       adsurl = {https://ui.adsabs.harvard.edu/abs/2024A&A...686L..12P}
}

@article{Reynolds2003,
   author = {Christopher S. Reynolds and Michael A. Nowak},
   doi = {10.1016/S0370-1573(02)00584-7},
   issn = {03701573},
   issue = {6},
   journal = {Physics Reports},
   month = {4},
   pages = {389-466},
   title = {Fluorescent iron lines as a probe of astrophysical black hole systems},
   volume = {377},
   url = {https://ui.adsabs.harvard.edu/abs/2003PhR...377..389R/abstract},
   year = {2003},
}

@ARTICLE{atri2020,
       author = {{Atri}, P. and {Miller-Jones}, J.~C.~A. and {Bahramian}, A. and {Plotkin}, R.~M. and {Deller}, A.~T. and {Jonker}, P.~G. and {Maccarone}, T.~J. and {Sivakoff}, G.~R. and {Soria}, R. and {Altamirano}, D. and {Belloni}, T. and {Fender}, R. and {Koerding}, E. and {Maitra}, D. and {Markoff}, S. and {Migliari}, S. and {Russell}, D. and {Russell}, T. and {Sarazin}, C.~L. and {Tetarenko}, A.~J. and {Tudose}, V.},
        title = "{A radio parallax to the black hole X-ray binary MAXI J1820+070}",
      journal = {\mnras},
         year = 2020,
        month = mar,
       volume = {493},
       number = {1},
        pages = {L81-L86},
          doi = {10.1093/mnrasl/slaa010},
archivePrefix = {arXiv},
       eprint = {1912.04525},
 primaryClass = {astro-ph.HE},
       adsurl = {https://ui.adsabs.harvard.edu/abs/2020MNRAS.493L..81A}
}

@ARTICLE{car,
       author = {{Carotenuto}, F. and {Corbel}, S. and {Tremou}, E. and {Russell}, T.~D. and {Tzioumis}, A. and {Fender}, R.~P. and {Woudt}, P.~A. and {Motta}, S.~E. and {Miller-Jones}, J.~C.~A. and {Chauhan}, J. and {Tetarenko}, A.~J. and {Sivakoff}, G.~R. and {Heywood}, I. and {Horesh}, A. and {van der Horst}, A.~J. and {Koerding}, E. and {Mooley}, K.~P.},
        title = "{The black hole transient MAXI J1348-630: evolution of the compact and transient jets during its 2019/2020 outburst}",
      journal = {\mnras},
         year = 2021,
        month = jun,
       volume = {504},
       number = {1},
        pages = {444-468},
          doi = {10.1093/mnras/stab864},
archivePrefix = {arXiv},
       eprint = {2103.12190},
 primaryClass = {astro-ph.HE},
       adsurl = {https://ui.adsabs.harvard.edu/abs/2021MNRAS.504..444C}
}

@ARTICLE{chauhan_2021,
       author = {{Chauhan}, J. and {Miller-Jones}, J.~C.~A. and {Raja}, W. and {Allison}, J.~R. and {Jacob}, P.~F.~L. and {Anderson}, G.~E. and {Carotenuto}, F. and {Corbel}, S. and {Fender}, R. and {Hotan}, A. and {Whiting}, M. and {Woudt}, P.~A. and {Koribalski}, B. and {Mahony}, E.},
        title = "{Measuring the distance to the black hole candidate X-ray binary MAXI J1348-630 using H I absorption}",
      journal = {\mnras},
         year = 2021,
        month = feb,
       volume = {501},
       number = {1},
        pages = {L60-L64},
          doi = {10.1093/mnrasl/slaa195},
archivePrefix = {arXiv},
       eprint = {2009.14419},
 primaryClass = {astro-ph.HE},
       adsurl = {https://ui.adsabs.harvard.edu/abs/2021MNRAS.501L..60C}
}

@ARTICLE{zhang,
       author = {{Zhang}, L. and {Altamirano}, D. and {C{\'u}neo}, V.~A. and {Alabarta}, K. and {Enoto}, T. and {Homan}, J. and {Remillard}, R.~A. and {Uttley}, P. and {Vincentelli}, F.~M. and {Arzoumanian}, Z. and {Bult}, P. and {Gendreau}, K.~C. and {Markwardt}, C. and {Sanna}, A. and {Strohmayer}, T.~E. and {Steiner}, J.~F. and {Basak}, A. and {Neilsen}, J. and {Tombesi}, F.},
        title = "{NICER observations reveal that the X-ray transient MAXI J1348-630 is a black hole X-ray binary}",
      journal = {\mnras},
         year = 2020,
        month = nov,
       volume = {499},
       number = {1},
        pages = {851-861},
          doi = {10.1093/mnras/staa2842},
archivePrefix = {arXiv},
       eprint = {2009.07749},
 primaryClass = {astro-ph.HE},
       adsurl = {https://ui.adsabs.harvard.edu/abs/2020MNRAS.499..851Z}
}

@misc{1820_isco,
    title={Continuum emission from within the plunging region of black hole discs},
    author={Andrew Mummery and Adam Ingram and Shane Davis and Andrew Fabian},
    year={2024},
    eprint={2405.09175},
    archivePrefix={arXiv},
    primaryClass={astro-ph.HE}
}

@article{Fabian_2020,
   title={The soft state of the black hole transient source MAXI J1820+070: emission from the edge of the plunge region?},
   volume={493},
   ISSN={1365-2966},
   url={http://dx.doi.org/10.1093/mnras/staa564},
   DOI={10.1093/mnras/staa564},
   number={4},
   journal={Monthly Notices of the Royal Astronomical Society},
   publisher={Oxford University Press (OUP)},
   author={Fabian, A C and Buisson, D J and Kosec, P and Reynolds, C S and Wilkins, D R and Tomsick, J A and Walton, D J and Gandhi, P and Altamirano, D and Arzoumanian, Z and Cackett, E M and Dyda, S and Garcia, J A and Gendreau, K C and Grefenstette, B W and Homan, J and Kara, E and Ludlam, R M and Miller, J M and Steiner, J F},
   year={2020},
   month=feb, pages={5389–5396} }

@article{Wood_2021,
   title={The varying kinematics of multiple ejecta from the black hole X-ray binary MAXI J1820 + 070},
   volume={505},
   ISSN={1365-2966},
   url={http://dx.doi.org/10.1093/mnras/stab1479},
   DOI={10.1093/mnras/stab1479},
   number={3},
   journal={Monthly Notices of the Royal Astronomical Society},
   publisher={Oxford University Press (OUP)},
   author={Wood, C M and Miller-Jones, J C A and Homan, J and Bright, J S and Motta, S E and Fender, R P and Markoff, S and Belloni, T M and Körding, E G and Maitra, D and Migliari, S and Russell, D M and Russell, T D and Sarazin, C L and Soria, R and Tetarenko, A J and Tudose, V},
   year={2021},
   month=may, pages={3393–3403} }

@ARTICLE{wilm,
       author = {{Wilms}, J. and {Allen}, A. and {McCray}, R.},
        title = "{On the Absorption of X-Rays in the Interstellar Medium}",
      journal = {\apj},
         year = 2000,
        month = oct,
       volume = {542},
       number = {2},
        pages = {914-924},
          doi = {10.1086/317016},
archivePrefix = {arXiv},
       eprint = {astro-ph/0008425},
 primaryClass = {astro-ph},
       adsurl = {https://ui.adsabs.harvard.edu/abs/2000ApJ...542..914W}
}

@article{Tao_2018,
   title={Swift observations of the bright uncatalogued X-ray transient MAXI J1535-571},
   volume={480},
   ISSN={1365-2966},
   url={http://dx.doi.org/10.1093/mnras/sty2157},
   DOI={10.1093/mnras/sty2157},
   number={4},
   journal={Monthly Notices of the Royal Astronomical Society},
   publisher={Oxford University Press (OUP)},
   author={Tao, Lian and Chen, YuPeng and GÜNGÖR, Can and Huang, Yue and Lu, FangJun and Qu, JinLu and Song, LiMing and Zhang, Liang and Zhang, Shu and Zhang, ShuangNan},
   year={2018},
   month=aug, pages={4443–4454} }

@ARTICLE{nakahira_2018,
       author = {{Nakahira}, Satoshi and {Shidatsu}, Megumi and {Makishima}, Kazuo and {Ueda}, Yoshihiro and {Yamaoka}, Kazutaka and {Mihara}, Tatehiro and {Negoro}, Hitoshi and {Kawase}, Tomofumi and {Kawai}, Nobuyuki and {Morita}, Kotaro},
        title = "{Discovery and state transitions of the new Galactic black hole candidate MAXI J1535-571}",
      journal = {\pasj},
         year = 2018,
        month = oct,
       volume = {70},
       number = {5},
          eid = {95},
        pages = {95},
          doi = {10.1093/pasj/psy093},
archivePrefix = {arXiv},
       eprint = {1804.00800},
 primaryClass = {astro-ph.HE},
       adsurl = {https://ui.adsabs.harvard.edu/abs/2018PASJ...70...95N}
}

@ARTICLE{Russell_2020,
       author = {{Russell}, T.~D. and {Lucchini}, M. and {Tetarenko}, A.~J. and {Miller-Jones}, J.~C.~A. and {Sivakoff}, G.~R. and {Krau{\ss}}, F. and {Mulaudzi}, W. and {Baglio}, M.~C. and {Russell}, D.~M. and {Altamirano}, D. and {Ceccobello}, C. and {Corbel}, S. and {Degenaar}, N. and {van den Eijnden}, J. and {Fender}, R. and {Heinz}, S. and {Koljonen}, K.~I.~I. and {Maitra}, D. and {Markoff}, S. and {Migliari}, S. and {Parikh}, A.~S. and {Plotkin}, R.~M. and {Rupen}, M. and {Sarazin}, C. and {Soria}, R. and {Wijnands}, R.},
        title = "{Rapid compact jet quenching in the Galactic black hole candidate X-ray binary MAXI J1535-571}",
      journal = {\mnras},
         year = 2020,
        month = nov,
       volume = {498},
       number = {4},
        pages = {5772-5785},
          doi = {10.1093/mnras/staa2650},
archivePrefix = {arXiv},
       eprint = {2008.11216},
 primaryClass = {astro-ph.HE},
       adsurl = {https://ui.adsabs.harvard.edu/abs/2020MNRAS.498.5772R}
}

@ARTICLE{fend04,
       author = {{Fender}, R.~P. and {Belloni}, T.~M. and {Gallo}, E.},
        title = "{Towards a unified model for black hole X-ray binary jets}",
      journal = {\mnras},
         year = 2004,
        month = dec,
       volume = {355},
       number = {4},
        pages = {1105-1118},
          doi = {10.1111/j.1365-2966.2004.08384.x},
archivePrefix = {arXiv},
       eprint = {astro-ph/0409360},
 primaryClass = {astro-ph},
       adsurl = {https://ui.adsabs.harvard.edu/abs/2004MNRAS.355.1105F}
}

@ARTICLE{kara19,
       author = {{Kara}, E. and {Steiner}, J.~F. and {Fabian}, A.~C. and {Cackett}, E.~M. and {Uttley}, P. and {Remillard}, R.~A. and {Gendreau}, K.~C. and {Arzoumanian}, Z. and {Altamirano}, D. and {Eikenberry}, S. and {Enoto}, T. and {Homan}, J. and {Neilsen}, J. and {Stevens}, A.~L.},
        title = "{The corona contracts in a black-hole transient}",
      journal = {\nat},
         year = 2019,
        month = jan,
       volume = {565},
       number = {7738},
        pages = {198-201},
          doi = {10.1038/s41586-018-0803-x},
archivePrefix = {arXiv},
       eprint = {1901.03877},
 primaryClass = {astro-ph.HE},
       adsurl = {https://ui.adsabs.harvard.edu/abs/2019Natur.565..198K}
}

@article{Shidatsu_2019,
doi = {10.3847/1538-4357/ab09ff},
url = {https://dx.doi.org/10.3847/1538-4357/ab09ff},
year = {2019},
month = {apr},
publisher = {The American Astronomical Society},
volume = {874},
number = {2},
pages = {183},
author = {Megumi Shidatsu and Satoshi Nakahira and Katsuhiro L. Murata and Ryo Adachi and Nobuyuki Kawai and Yoshihiro Ueda and Hitoshi Negoro},
title = {X-Ray and Optical Monitoring of State Transitions in MAXI J1820+070},
journal = {The Astrophysical Journal}
}

@article{Russell_2019,
doi = {10.3847/1538-4357/ab3d36},
url = {https://dx.doi.org/10.3847/1538-4357/ab3d36},
year = {2019},
month = {oct},
publisher = {The American Astronomical Society},
volume = {883},
number = {2},
pages = {198},
author = {T. D. Russell and A. J. Tetarenko and J. C. A. Miller-Jones and G. R. Sivakoff and A. S. Parikh and S. Rapisarda and R. Wijnands and S. Corbel and E. Tremou and D. Altamirano and M. C. Baglio and C. Ceccobello and N. Degenaar and J. van den Eijnden and R. Fender and I. Heywood and H. A. Krimm and M. Lucchini and S. Markoff and D. M. Russell and R. Soria and P. A. Woudt},
title = {Disk–Jet Coupling in the 2017/2018 Outburst of the Galactic Black Hole Candidate X-Ray Binary MAXI J1535–571},
journal = {The Astrophysical Journal}
}

@ARTICLE{cao2022,
       author = {{Cao}, Zheng and {Lucchini}, Matteo and {Markoff}, Sera and {Connors}, Riley M.~T. and {Grinberg}, Victoria},
        title = "{Evidence for an expanding corona based on spectral-timing modelling of multiple black hole X-ray binaries}",
      journal = {\mnras},
         year = 2022,
        month = jan,
       volume = {509},
       number = {2},
        pages = {2517-2531},
          doi = {10.1093/mnras/stab3080},
archivePrefix = {arXiv},
       eprint = {2110.10547},
 primaryClass = {astro-ph.HE},
       adsurl = {https://ui.adsabs.harvard.edu/abs/2022MNRAS.509.2517C}
}

@article{zhang2023,
   title={ANICERlook at the jet-like corona of MAXI J1535−571 through type-B quasi-periodic oscillations},
   volume={520},
   ISSN={1365-2966},
   url={http://dx.doi.org/10.1093/mnras/stad460},
   DOI={10.1093/mnras/stad460},
   number={4},
   journal={Monthly Notices of the Royal Astronomical Society},
   publisher={Oxford University Press (OUP)},
   author={Zhang, Yuexin and Méndez, Mariano and García, Federico and Altamirano, Diego and Belloni, Tomaso M and Alabarta, Kevin and Zhang, Liang and Bellavita, Candela and Rawat, Divya and Ma, Ruican},
   year={2023},
   month=feb, pages={5144–5156} 
}

@ARTICLE{gilles1991,
       author = {{Henri}, Gilles and {Pelletier}, Guy},
        title = "{Relativistic Electron-Positron Beam Formation in the Framework of the Two-Flow Model for Active Galactic Nuclei}",
      journal = {\apjl},
         year = 1991,
        month = dec,
       volume = {383},
        pages = {L7},
          doi = {10.1086/186228},
       adsurl = {https://ui.adsabs.harvard.edu/abs/1991ApJ...383L...7H}
}

@ARTICLE{you2021,
       author = {{You}, Bei and {Tuo}, Yuoli and {Li}, Chengzhe and {Wang}, Wei and {Zhang}, Shuang-Nan and {Zhang}, Shu and {Ge}, Mingyu and {Luo}, Chong and {Liu}, Bifang and {Yuan}, Weimin and {Dai}, Zigao and {Liu}, Jifeng and {Qiao}, Erlin and {Jin}, Chichuan and {Liu}, Zhu and {Czerny}, Bozena and {Wu}, Qingwen and {Bu}, Qingcui and {Cai}, Ce and {Cao}, Xuelei and {Chang}, Zhi and {Chen}, Gang and {Chen}, Li and {Chen}, Tianxiang and {Chen}, Yibao and {Chen}, Yong and {Chen}, Yupeng and {Cui}, Wei and {Cui}, Weiwei and {Deng}, Jingkang and {Dong}, Yongwei and {Du}, Yuanyuan and {Fu}, Minxue and {Gao}, Guanhua and {Gao}, He and {Gao}, Min and {Gu}, Yudong and {Guan}, Ju and {Guo}, Chengcheng and {Han}, Dawei and {Huang}, Yue and {Huo}, Jia and {Jia}, Shumei and {Jiang}, Luhua and {Jiang}, Weichun and {Jin}, Jing and {Jin}, Yongjie and {Kong}, Lingda and {Li}, Bing and {Li}, Chengkui and {Li}, Gang and {Li}, Maoshun and {Li}, Tipei and {Li}, Wei and {Li}, Xian and {Li}, Xiaobo and {Li}, Xufang and {Li}, Yanguo and {Li}, Zhengwei and {Liang}, Xiaohua and {Liao}, Jinyuan and {Liu}, Congzhan and {Liu}, Guoqing and {Liu}, Hongwei and {Liu}, Xiaojing and {Liu}, Yinong and {Lu}, Bo and {Lu}, Fangjun and {Lu}, Xuefeng and {Luo}, Qi and {Luo}, Tao and {Ma}, Xiang and {Meng}, Bin and {Nang}, Yi and {Nie}, Jianyin and {Ou}, Ge and {Qu}, Jinlu and {Sai}, Na and {Shang}, Rencheng and {Song}, Liming and {Song}, Xinying and {Sun}, Liang and {Tan}, Ying and {Tao}, Lian and {Wang}, Chen and {Wang}, Guofeng and {Wang}, Juan and {Wang}, Lingjun and {Wang}, Wenshuai and {Wang}, Yusa and {Wen}, Xiangyang and {Wu}, Baiyang and {Wu}, Bobing and {Wu}, Mei and {Xiao}, Guangcheng and {Xiao}, Shuo and {Xiong}, Shaolin and {Xu}, Yupeng and {Yang}, Jiawei and {Yang}, Sheng and {Yang}, Yanji and {Yi}, Qibin and {Yin}, Qianqing and {You}, Yuan and {Zhang}, Aimei and {Zhang}, Chengmo and {Zhang}, Fan and {Zhang}, Hongmei and {Zhang}, Juan and {Zhang}, Tong and {Zhang}, Wanchang and {Zhang}, Wei and {Zhang}, Wenzhao and {Zhang}, Yi and {Zhang}, Yifei and {Zhang}, Yongjie and {Zhang}, Yue and {Zhang}, Zhao and {Zhang}, Ziliang and {Zhao}, Haisheng and {Zhao}, Xiaofan and {Zheng}, Shijie and {Zhou}, Dengke and {Zhou}, Jianfeng and {Zhu}, Yuxuan and {Zhu}, Yue},
        title = "{Insight-HXMT observations of jet-like corona in a black hole X-ray binary MAXI J1820+070}",
      journal = {Nature Communications},
         year = 2021,
        month = jan,
       volume = {12},
          eid = {1025},
        pages = {1025},
          doi = {10.1038/s41467-021-21169-5},
archivePrefix = {arXiv},
       eprint = {2102.07602},
 primaryClass = {astro-ph.HE},
       adsurl = {https://ui.adsabs.harvard.edu/abs/2021NatCo..12.1025Y}
}

@INPROCEEDINGS{xspec,
       author = {{Arnaud}, K.~A.},
        title = "{XSPEC: The First Ten Years}",
    booktitle = {Astronomical Data Analysis Software and Systems V},
         year = 1996,
       editor = {{Jacoby}, George H. and {Barnes}, Jeannette},
       series = {Astronomical Society of the Pacific Conference Series},
       volume = {101},
        month = jan,
        pages = {17},
       adsurl = {https://ui.adsabs.harvard.edu/abs/1996ASPC..101...17A}
}

@ARTICLE{diskbb,
       author = {{Mitsuda}, K. and {Inoue}, H. and {Koyama}, K. and {Makishima}, K. and {Matsuoka}, M. and {Ogawara}, Y. and {Shibazaki}, N. and {Suzuki}, K. and {Tanaka}, Y. and {Hirano}, T.},
        title = "{Energy spectra of low-mass binary X-ray sources observed from Tenma.}",
      journal = {\pasj},
         year = 1984,
        month = jan,
       volume = {36},
        pages = {741-759},
       adsurl = {https://ui.adsabs.harvard.edu/abs/1984PASJ...36..741M}
}

@ARTICLE{tbabs,
       author = {{Wilms}, J. and {Allen}, A. and {McCray}, R.},
        title = "{On the Absorption of X-Rays in the Interstellar Medium}",
      journal = {\apj},
         year = 2000,
        month = oct,
       volume = {542},
       number = {2},
        pages = {914-924},
          doi = {10.1086/317016},
archivePrefix = {arXiv},
       eprint = {astro-ph/0008425},
 primaryClass = {astro-ph},
       adsurl = {https://ui.adsabs.harvard.edu/abs/2000ApJ...542..914W}
}

@ARTICLE{haardt91,
       author = {{Haardt}, F. and {Maraschi}, L.},
        title = "{A Two-Phase Model for the X-Ray Emission from Seyfert Galaxies}",
      journal = {\apjl},
         year = 1991,
        month = oct,
       volume = {380},
        pages = {L51},
          doi = {10.1086/186171},
       adsurl = {https://ui.adsabs.harvard.edu/abs/1991ApJ...380L..51H}
}

@article{Echibur_Trujillo_2024,
   title={Chasing the Break: Tracing the Full Evolution of a Black Hole X-Ray Binary Jet with Multiwavelength Spectral Modeling},
   volume={962},
   ISSN={1538-4357},
   url={http://dx.doi.org/10.3847/1538-4357/ad1a10},
   DOI={10.3847/1538-4357/ad1a10},
   number={2},
   journal={The Astrophysical Journal},
   publisher={American Astronomical Society},
   author={Echiburú-Trujillo, Constanza and Tetarenko, Alexandra J. and Haggard, Daryl and Russell, Thomas D. and Koljonen, Karri I. I. and Bahramian, Arash and Wang, Jingyi and Bremer, Michael and Bright, Joe and Casella, Piergiorgio and Russell, David M. and Altamirano, Diego and Baglio, M. Cristina and Belloni, Tomaso and Ceccobello, Chiara and Corbel, Stephane and Diaz Trigo, Maria and Maitra, Dipankar and Gabuya, Aldrin and Gallo, Elena and Heinz, Sebastian and Homan, Jeroen and Kara, Erin and Körding, Elmar and Lewis, Fraser and Lucchini, Matteo and Markoff, Sera and Migliari, Simone and Miller-Jones, James C. A. and Rodriguez, Jerome and Saikia, Payaswini and Sarazin, Craig L. and Shahbaz, Tariq and Sivakoff, Gregory and Soria, Roberto and Testa, Vincenzo and Tetarenko, Bailey E. and Tudose, Valeriu},
   year={2024},
   month=feb, pages={116} }

@MANUAL{relxill,
       author = {{Dauser}, T. and {Garcia}, J.},
        title = "{Relativistic Reflection with the Relxill model version 2.0}",
         year = 2022,
        month = jun,
}

@ARTICLE{dauser,
       author = {{Dauser}, T. and {Garcia}, J. and {Wilms}, J. and {B{\"o}ck}, M. and {Brenneman}, L.~W. and {Falanga}, M. and {Fukumura}, K. and {Reynolds}, C.~S.},
        title = "{Irradiation of an accretion disc by a jet: general properties and implications for spin measurements of black holes}",
      journal = {\mnras},
         year = 2013,
        month = apr,
       volume = {430},
       number = {3},
        pages = {1694-1708},
          doi = {10.1093/mnras/sts71010.48550/arXiv.1301.4922},
archivePrefix = {arXiv},
       eprint = {1301.4922},
 primaryClass = {astro-ph.HE},
       adsurl = {https://ui.adsabs.harvard.edu/abs/2013MNRAS.430.1694D}
}

@misc{nicer_pileup,
author = {NASA Goddard Space Flight Center},
title = {NICER Calibration Recommendation},
howpublished = {\url{https://heasarc.gsfc.nasa.gov/docs/nicer/analysis_threads/cal-recommend/}},
note = {10/1/2024},
year = {09/25/2024}
}

@article{Wang_2020hh,
   title={The Evolution of the Broadband Temporal Features Observed in the Black-hole Transient MAXI J1820+070 with Insight-HXMT},
   volume={896},
   ISSN={1538-4357},
   url={http://dx.doi.org/10.3847/1538-4357/ab8db4},
   DOI={10.3847/1538-4357/ab8db4},
   number={1},
   journal={The Astrophysical Journal},
   publisher={American Astronomical Society},
   author={Wang, Yanan and Ji, Long and Zhang, S. N. and Méndez, Mariano and Qu, J. L. and Maggi, Pierre and Ge, M. Y. and Qiao, Erlin and Tao, L. and Zhang, S. and Altamirano, Diego and Zhang, L. and Ma, X. and Lu, F. J. and Li, T. P. and Huang, Y. and Zheng, S. J. and Chen, Y. P. and Chang, Z. and Tuo, Y. L. and Güngör, C. and Song, L. M. and Xu, Y. P. and Cao, X. L. and Chen, Y. and Liu, C. Z. and Bu, Q. C. and Cai, C. and Chen, G. and Chen, L. and Chen, T. X. and Chen, Y. B. and Cui, W. and Cui, W. W. and Deng, J. K. and Dong, Y. W. and Du, Y. Y. and Fu, M. X. and Gao, G. H. and Gao, H. and Gao, M. and Gu, Y. D. and Guan, J. and Guo, C. C. and Han, D. W. and Huo, J. and Jia, S. M. and Jiang, L. H. and Jiang, W. C. and Jin, J. and Jin, Y. J. and Kong, L. D. and Li, B. and Li, C. K. and Li, G. and Li, M. S. and Li, W. and Li, X. and Li, X. B. and Li, X. F. and Li, Y. G. and Li, Z. W. and Liang, X. H. and Liao, J. Y. and Liu, G. Q. and Liu, H. W. and Liu, X. J. and Liu, Y. N. and Lu, B. and Lu, X. F. and Luo, Q. and Luo, T. and Meng, B. and Nang, Y. and Nie, J. Y. and Ou, G. and Sai, N. and Shang, R. C. and Song, X. Y. and Sun, L. and Tan, Y. and Wang, G. F. and Wang, J. and Wang, W. S. and Wang, Y. D. and Wang, Y. S. and Wen, X. Y. and Wu, B. B. and Wu, B. Y. and Wu, M. and Xiao, G. C. and Xiao, S. and Xiong, S. L. and Yang, J. W. and Yang, S. and Yang, Y. J. and Yang, Y. J. and Yi, Q. B. and Yin, Q. Q. and You, Y. and Zhang, A. M. and Zhang, C. M. and Zhang, F. and Zhang, H. M. and Zhang, J. and Zhang, T. and Zhang, W. C. and Zhang, W. and Zhang, W. Z. and Zhang, Y. and Zhang, Y. F. and Zhang, Y. J. and Zhang, Y. and Zhang, Z. and Zhang, Z. and Zhang, Z. L. and Zhao, H. S. and Zhao, X. F. and Zhou, D. K. and Zhou, J. F. and Zhuang, R. L. and Zhu, Y. X. and Zhu, Y. and Wang, Lingjun},
   year={2020},
   month=jun, pages={33} }

@article{Veledina_2023_ixpe,
doi = {10.3847/2041-8213/ad0781},
url = {https://dx.doi.org/10.3847/2041-8213/ad0781},
year = {2023},
month = {nov},
publisher = {The American Astronomical Society},
volume = {958},
number = {1},
pages = {L16},
author = {Alexandra Veledina and Fabio Muleri and Michal Dovčiak and Juri Poutanen and Ajay Ratheesh and Fiamma Capitanio and Giorgio Matt and Paolo Soffitta and Allyn F. Tennant and Michela Negro and Philip Kaaret and Enrico Costa and Adam Ingram and Jiří Svoboda and Henric Krawczynski and Stefano Bianchi and James F. Steiner and Javier A. García and Vadim Kravtsov and Anagha P. Nitindala and Melissa Ewing and Guglielmo Mastroserio and Andrea Marinucci and Francesco Ursini and Francesco Tombesi and Sergey S. Tsygankov and Yi-Jung Yang and Martin C. Weisskopf and Sergei A. Trushkin and Elise Egron and Maria Noemi Iacolina and Maura Pilia and Lorenzo Marra and Romana Mikušincová and Edward Nathan and Maxime Parra and Pierre-Olivier Petrucci and Jakub Podgorný and Stefano Tugliani and Silvia Zane and Wenda Zhang and Iván Agudo and Lucio A. Antonelli and Matteo Bachetti and Luca Baldini and Wayne H. Baumgartner and Ronaldo Bellazzini and Stephen D. Bongiorno and Raffaella Bonino and Alessandro Brez and Niccolò Bucciantini and Simone Castellano and Elisabetta Cavazzuti and Chien-Ting Chen and Stefano Ciprini and Alessandra De Rosa and Ettore Del Monte and Laura Di Gesu and Niccolò Di Lalla and Alessandro Di Marco and Immacolata Donnarumma and Victor Doroshenko and Steven R. Ehlert and Teruaki Enoto and Yuri Evangelista and Sergio Fabiani and Riccardo Ferrazzoli and Shuichi Gunji and Kiyoshi Hayashida and Jeremy Heyl and Wataru Iwakiri and Svetlana G. Jorstad and Vladimir Karas and Fabian Kislat and Takao Kitaguchi and Jeffery J. Kolodziejczak and Fabio La Monaca and Luca Latronico and Ioannis Liodakis and Simone Maldera and Alberto Manfreda and Frédéric Marin and Alan P. Marscher and Herman L. Marshall and Francesco Massaro and Ikuyuki Mitsuishi and Tsunefumi Mizuno and Chi-Yung Ng and Stephen L. O’Dell and Nicola Omodei and Chiara Oppedisano and Alessandro Papitto and George G. Pavlov and Abel L. Peirson and Matteo Perri and Melissa Pesce-Rollins and Andrea Possenti and Simonetta Puccetti and Brian D. Ramsey and John Rankin and Oliver J. Roberts and Roger W. Romani and Carmelo Sgrò and Patrick Slane and Gloria Spandre and Douglas A. Swartz and Toru Tamagawa and Fabrizio Tavecchio and Roberto Taverna and Yuzuru Tawara and Nicholas E. Thomas and Alessio Trois and Roberto Turolla and Jacco Vink and Kinwah Wu and Fei Xie},
title = {Discovery of X-Ray Polarization from the Black Hole Transient Swift J1727.8−1613},
journal = {The Astrophysical Journal Letters}
}

@ARTICLE{cyg_ixpe,
       author = {{Krawczynski}, Henric and {Muleri}, Fabio and {Dov{\v{c}}iak}, Michal and {Veledina}, Alexandra and {Rodriguez Cavero}, Nicole and {Svoboda}, Jiri and {Ingram}, Adam and {Matt}, Giorgio and {Garcia}, Javier A. and {Loktev}, Vladislav and {Negro}, Michela and {Poutanen}, Juri and {Kitaguchi}, Takao and {Podgorn{\'y}}, Jakub and {Rankin}, John and {Zhang}, Wenda and {Berdyugin}, Andrei and {Berdyugina}, Svetlana V. and {Bianchi}, Stefano and {Blinov}, Dmitry and {Capitanio}, Fiamma and {Di Lalla}, Niccol{\`o} and {Draghis}, Paul and {Fabiani}, Sergio and {Kagitani}, Masato and {Kravtsov}, Vadim and {Kiehlmann}, Sebastian and {Latronico}, Luca and {Lutovinov}, Alexander A. and {Mandarakas}, Nikos and {Marin}, Fr{\'e}d{\'e}ric and {Marinucci}, Andrea and {Miller}, Jon M. and {Mizuno}, Tsunefumi and {Molkov}, Sergey V. and {Omodei}, Nicola and {Petrucci}, Pierre-Olivier and {Ratheesh}, Ajay and {Sakanoi}, Takeshi and {Semena}, Andrei N. and {Skalidis}, Raphael and {Soffitta}, Paolo and {Tennant}, Allyn F. and {Thalhammer}, Phillipp and {Tombesi}, Francesco and {Weisskopf}, Martin C. and {Wilms}, Joern and {Zhang}, Sixuan and {Agudo}, Iv{\'a}n and {Antonelli}, Lucio A. and {Bachetti}, Matteo and {Baldini}, Luca and {Baumgartner}, Wayne H. and {Bellazzini}, Ronaldo and {Bongiorno}, Stephen D. and {Bonino}, Raffaella and {Brez}, Alessandro and {Bucciantini}, Niccol{\`o} and {Castellano}, Simone and {Cavazzuti}, Elisabetta and {Ciprini}, Stefano and {Costa}, Enrico and {De Rosa}, Alessandra and {Del Monte}, Ettore and {Di Gesu}, Laura and {Di Marco}, Alessandro and {Donnarumma}, Immacolata and {Doroshenko}, Victor and {Ehlert}, Steven R. and {Enoto}, Teruaki and {Evangelista}, Yuri and {Ferrazzoli}, Riccardo and {Gunji}, Shuichi and {Hayashida}, Kiyoshi and {Heyl}, Jeremy and {Iwakiri}, Wataru and {Jorstad}, Svetlana G. and {Karas}, Vladimir and {Kolodziejczak}, Jeffery J. and {La Monaca}, Fabio and {Liodakis}, Ioannis and {Maldera}, Simone and {Manfreda}, Alberto and {Marscher}, Alan P. and {Marshall}, Herman L. and {Mitsuishi}, Ikuyuki and {Ng}, Chi-Yung and {O{\textquoteright}Dell}, Stephen L. and {Oppedisano}, Chiara and {Papitto}, Alessandro and {Pavlov}, George G. and {Peirson}, Abel L. and {Perri}, Matteo and {Pesce-Rollins}, Melissa and {Pilia}, Maura and {Possenti}, Andrea and {Puccetti}, Simonetta and {Ramsey}, Brian D. and {Romani}, Roger W. and {Sgr{\`o}}, Carmelo and {Slane}, Patrick and {Spandre}, Gloria and {Tamagawa}, Toru and {Tavecchio}, Fabrizio and {Taverna}, Roberto and {Tawara}, Yuzuru and {Thomas}, Nicholas E. and {Trois}, Alessio and {Tsygankov}, Sergey and {Turolla}, Roberto and {Vink}, Jacco and {Wu}, Kinwah and {Xie}, Fei and {Zane}, Silvia},
        title = "{Polarized x-rays constrain the disk-jet geometry in the black hole x-ray binary Cygnus X-1}",
      journal = {Science},
         year = 2022,
        month = nov,
       volume = {378},
       number = {6620},
        pages = {650-654},
          doi = {10.1126/science.add5399},
archivePrefix = {arXiv},
       eprint = {2206.09972},
 primaryClass = {astro-ph.HE},
       adsurl = {https://ui.adsabs.harvard.edu/abs/2022Sci...378..650K}
}

@misc{liao2024,
      title={Tracking the jet-like corona of black hole Swift J1727.8-1613 during a flare state through Type-C quasi-periodic oscillations}, 
      author={Jie Liao and Ning Chang and Lang Cui and Pengfei Jiang and Didong Mou and Yongfeng Huang and Tao An and Luis C. Ho and Hua Feng and Yu-Cong Fu and Hongmin Cao and Xiang Liu},
      year={2024},
      eprint={2410.06574},
      archivePrefix={arXiv},
      primaryClass={astro-ph.HE},
      url={https://arxiv.org/abs/2410.06574}, 
}

@misc{sripada2022,
      title={Type-B QPOs in the black hole source H1743-322 and their association with Comptonization region and Jet}, 
      author={Harikrishna Sripada and Sriram Kandulapati},
      year={2022},
      eprint={2209.01643},
      archivePrefix={arXiv},
      primaryClass={astro-ph.HE},
      url={https://arxiv.org/abs/2209.01643}, 
}

@article{Liu_2022,
doi = {10.3847/1538-4357/ac88c6},
url = {https://dx.doi.org/10.3847/1538-4357/ac88c6},
year = {2022},
month = {oct},
publisher = {The American Astronomical Society},
volume = {938},
number = {2},
pages = {108},
author = {H. X. Liu and Y. Huang and Q. C. Bu and W. Yu and Z. X. Yang and L. Zhang and L. D. Kong and G. C. Xiao and J. L. Qu and S. N. Zhang and S. Zhang and L. M. Song and S. M. Jia and X. Ma and L. Tao and M. Y. Ge and Q. Z. Liu and J. Z. Yan and R. C. Ma and X. Q. Ren and D. K. Zhou and T. M. Li and B. Y. Wu and Y. C. Xu and Y. F. Du and Y. C. Fu and Y. X. Xiao and G. Q. Ding and X. X. Yu},
title = {Transitions and Origin of the Type-B Quasi-periodic Oscillations in the Black Hole X-Ray Binary MAXI J1348–630},
journal = {The Astrophysical Journal}
}

@article{Garc_a_2020,
   title={A two-component Comptonization model for the type-B QPO in MAXI J1348−630},
   volume={501},
   ISSN={1365-2966},
   url={http://dx.doi.org/10.1093/mnras/staa3944},
   DOI={10.1093/mnras/staa3944},
   number={3},
   journal={Monthly Notices of the Royal Astronomical Society},
   publisher={Oxford University Press (OUP)},
   author={García, Federico and Méndez, Mariano and Karpouzas, Konstantinos and Belloni, Tomaso and Zhang, Liang and Altamirano, Diego},
   year={2020},
   month=dec, pages={3173–3182} }

@ARTICLE{Garcia2022,
       author = {{Garc{\'\i}a}, Federico and {Karpouzas}, Konstantinos and {M{\'e}ndez}, Mariano and {Zhang}, Liang and {Zhang}, Yuexin and {Belloni}, Tomaso and {Altamirano}, Diego},
        title = "{The evolving properties of the corona of GRS 1915+105: a spectral-timing perspective through variable-Comptonization modelling}",
      journal = {\mnras},
         year = 2022,
        month = jul,
       volume = {513},
       number = {3},
        pages = {4196-4207},
          doi = {10.1093/mnras/stac1202},
archivePrefix = {arXiv},
       eprint = {2204.13279},
 primaryClass = {astro-ph.HE},
       adsurl = {https://ui.adsabs.harvard.edu/abs/2022MNRAS.513.4196G}
}

@ARTICLE{Remillard2006,
       author = {{Remillard}, Ronald A. and {McClintock}, Jeffrey E.},
        title = "{X-Ray Properties of Black-Hole Binaries}",
      journal = {\araa},
         year = "2006",
        month = "Sep",
       volume = {44},
       number = {1},
        pages = {49-92},
          doi = {10.1146/annurev.astro.44.051905.092532},
archivePrefix = {arXiv},
       eprint = {astro-ph/0606352},
 primaryClass = {astro-ph},
       adsurl = {https://ui.adsabs.harvard.edu/abs/2006ARA&A..44...49R}
}

@INBOOK{Belloni2010,
       author = {{Belloni}, T.~M.},
        title = "{States and Transitions in Black Hole Binaries}",
    booktitle = {The Jet Paradigm, Lecture Notes in Physics, Volume 794. ISBN 978-3-540-76936-1. Springer-Verlag Berlin Heidelberg, 2010, p. 53},
         year = "2010",
       editor = {{Belloni}, Tomaso},
       volume = {794},
        pages = {53},
          doi = {10.1007/978-3-540-76937-8_3},
       adsurl = {https://ui.adsabs.harvard.edu/abs/2010LNP...794...53B}
}

@INBOOK{Fender2006,
       author = {{Fender}, Rob},
        title = "{Jets from X-ray binaries}",
    booktitle = {In: Compact stellar X-ray sources. Edited by Walter Lewin \&amp; Michiel van der Klis. Cambridge Astrophysics Series, No. 39. Cambridge, UK: Cambridge University Press, ISBN 978-0-521-82659-4, ISBN 0-521-82659-4, DOI: 10.2277/0521826594, 2006, p. 381 - 419},
         year = "2006",
       volume = {39},
        pages = {381-419},
       adsurl = {https://ui.adsabs.harvard.edu/abs/2006csxs.book..381F}
}

@ARTICLE{Fender2004,
       author = {{Fender}, R.~P. and {Belloni}, T.~M. and {Gallo}, E.},
        title = "{Towards a unified model for black hole X-ray binary jets}",
      journal = {\mnras},
         year = "2004",
        month = "Dec",
       volume = {355},
       number = {4},
        pages = {1105-1118},
          doi = {10.1111/j.1365-2966.2004.08384.x},
archivePrefix = {arXiv},
       eprint = {astro-ph/0409360},
 primaryClass = {astro-ph},
       adsurl = {https://ui.adsabs.harvard.edu/abs/2004MNRAS.355.1105F}
}

@ARTICLE{Vadawale2003,
       author = {{Vadawale}, S.~V. and {Rao}, A.~R. and {Naik}, S. and {Yadav}, J.~S. and {Ishwara-Chandra}, C.~H. and {Pramesh Rao}, A. and {Pooley}, G.~G.},
        title = "{On the Origin of the Various Types of Radio Emission in GRS 1915+105}",
      journal = {\apj},
         year = 2003,
        month = nov,
       volume = {597},
       number = {2},
        pages = {1023-1035},
          doi = {10.1086/378672},
archivePrefix = {arXiv},
       eprint = {astro-ph/0308096},
 primaryClass = {astro-ph},
       adsurl = {https://ui.adsabs.harvard.edu/abs/2003ApJ...597.1023V}
}

@ARTICLE{Hjellming1988,
       author = {{Hjellming}, R.~M. and {Johnston}, K.~J.},
        title = "{Radio Emission from Conical Jets Associated with X-Ray Binaries}",
      journal = {\apj},
         year = 1988,
        month = may,
       volume = {328},
        pages = {600},
          doi = {10.1086/166318},
       adsurl = {https://ui.adsabs.harvard.edu/abs/1988ApJ...328..600H}
}

@ARTICLE{Fender2001,
       author = {{Fender}, R.~P.},
        title = "{Powerful jets from black hole X-ray binaries in low/hard X-ray states}",
      journal = {\mnras},
         year = "2001",
        month = "Mar",
       volume = {322},
       number = {1},
        pages = {31-42},
          doi = {10.1046/j.1365-8711.2001.04080.x},
archivePrefix = {arXiv},
       eprint = {astro-ph/0008447},
 primaryClass = {astro-ph},
       adsurl = {https://ui.adsabs.harvard.edu/abs/2001MNRAS.322...31F}
}

@ARTICLE{Harmon1995,
       author = {{Harmon}, B.~A. and {Wilson}, C.~A. and {Zhang}, S.~N. and {Paciesas}, W.~S. and {Fishman}, G.~J. and {Hjellming}, R.~M. and {Rupen}, M.~P. and {Scott}, D.~M. and {Briggs}, M.~S. and {Rubin}, B.~C.},
        title = "{Correlations between X-ray outbursts and relativistic ejections in the X-ray transient GRO J1655 - 40}",
      journal = {\nat},
         year = 1995,
        month = apr,
       volume = {374},
       number = {6524},
        pages = {703-706},
          doi = {10.1038/374703a0},
       adsurl = {https://ui.adsabs.harvard.edu/abs/1995Natur.374..703H}
}

@ARTICLE{Ross1993,
       author = {{Ross}, R.~R. and {Fabian}, A.~C.},
        title = "{The effects of photoionization on X-ray reflection spectra in active galactic nuclei.}",
      journal = {\mnras},
         year = 1993,
        month = mar,
       volume = {261},
        pages = {74-82},
          doi = {10.1093/mnras/261.1.74},
       adsurl = {https://ui.adsabs.harvard.edu/abs/1993MNRAS.261...74R}
}

@ARTICLE{Bharali2019,
       author = {{Bharali}, Priya and {Chandra}, Sunil and {Chauhan}, Jaiverdhan and {Garc{\'\i}a}, Javier A. and {Roy}, Jayashree and {Boettcher}, Markus and {Boruah}, Kalyanee},
        title = "{Re-awakening of GRS 1716-249 after 23 yr, observed by Swift/XRT and NuSTAR}",
      journal = {\mnras},
         year = 2019,
        month = aug,
       volume = {487},
       number = {3},
        pages = {3150-3161},
          doi = {10.1093/mnras/stz1492},
archivePrefix = {arXiv},
       eprint = {1905.12657},
 primaryClass = {astro-ph.HE},
       adsurl = {https://ui.adsabs.harvard.edu/abs/2019MNRAS.487.3150B}
}

@ARTICLE{Lightman1988,
       author = {{Lightman}, Alan P. and {White}, Timothy R.},
        title = "{Effects of Cold Matter in Active Galactic Nuclei: A Broad Hump in the X-Ray Spectra}",
      journal = {\apj},
         year = 1988,
        month = dec,
       volume = {335},
        pages = {57},
          doi = {10.1086/166905},
       adsurl = {https://ui.adsabs.harvard.edu/abs/1988ApJ...335...57L}
}

@ARTICLE{Haardt1993,
       author = {{Haardt}, Francesco and {Maraschi}, Laura},
        title = "{X-Ray Spectra from Two-Phase Accretion Disks}",
      journal = {\apj},
         year = 1993,
        month = aug,
       volume = {413},
        pages = {507},
          doi = {10.1086/173020},
       adsurl = {https://ui.adsabs.harvard.edu/abs/1993ApJ...413..507H}
}

@ARTICLE{Russell2015,
       author = {{Russell}, T.~D. and {Miller-Jones}, J.~C.~A. and {Curran}, P.~A. and {Soria}, R. and {Altamirano}, D. and {Corbel}, S. and {Coriat}, M. and {Moin}, A. and {Russell}, D.~M. and {Sivakoff}, G.~R. and {Slaven-Blair}, T.~J. and {Belloni}, T.~M. and {Fender}, R.~P. and {Heinz}, S. and {Jonker}, P.~G. and {Krimm}, H.~A. and {K{\"o}rding}, E.~G. and {Maitra}, D. and {Markoff}, S. and {Middleton}, M. and {Migliari}, S. and {Remillard}, R.~A. and {Rupen}, M.~P. and {Sarazin}, C.~L. and {Tetarenko}, A.~J. and {Torres}, M.~A.~P. and {Tudose}, V. and {Tzioumis}, A.~K.},
        title = "{Radio monitoring of the hard state jets in the 2011 outburst of MAXI J1836-194}",
      journal = {\mnras},
         year = 2015,
        month = jun,
       volume = {450},
       number = {2},
        pages = {1745-1759},
          doi = {10.1093/mnras/stv723},
archivePrefix = {arXiv},
       eprint = {1503.08634},
 primaryClass = {astro-ph.HE},
       adsurl = {https://ui.adsabs.harvard.edu/abs/2015MNRAS.450.1745R}
}

@ARTICLE{Kara2016,
       author = {{Kara}, E. and {Alston}, W.~N. and {Fabian}, A.~C. and {Cackett}, E.~M. and {Uttley}, P. and {Reynolds}, C.~S. and {Zoghbi}, A.},
        title = "{A global look at X-ray time lags in Seyfert galaxies}",
      journal = {\mnras},
         year = 2016,
        month = oct,
       volume = {462},
       number = {1},
        pages = {511-531},
          doi = {10.1093/mnras/stw1695},
archivePrefix = {arXiv},
       eprint = {1605.02631},
 primaryClass = {astro-ph.HE},
       adsurl = {https://ui.adsabs.harvard.edu/abs/2016MNRAS.462..511K}
}

@ARTICLE{Romano2006,
       author = {{Romano}, P. and {Campana}, S. and {Chincarini}, G. and {Cummings}, J. and {Cusumano}, G. and {Holland}, S.~T. and {Mangano}, V. and {Mineo}, T. and {Page}, K.~L. and {Pal'Shin}, V. and {Rol}, E. and {Sakamoto}, T. and {Zhang}, B. and {Aptekar}, R. and {Barbier}, S. and {Barthelmy}, S. and {Beardmore}, A.~P. and {Boyd}, P. and {Burrows}, D.~N. and {Capalbi}, M. and {Fenimore}, E.~E. and {Frederiks}, D. and {Gehrels}, N. and {Giommi}, P. and {Goad}, M.~R. and {Godet}, O. and {Golenetskii}, S. and {Guetta}, D. and {Kennea}, J.~A. and {La Parola}, V. and {Malesani}, D. and {Marshall}, F. and {Moretti}, A. and {Nousek}, J.~A. and {O'Brien}, P.~T. and {Osborne}, J.~P. and {Perri}, M. and {Tagliaferri}, G.},
        title = "{Panchromatic study of GRB 060124: from precursor to afterglow}",
      journal = {\aap},
         year = 2006,
        month = sep,
       volume = {456},
       number = {3},
        pages = {917-927},
          doi = {10.1051/0004-6361:20065071},
archivePrefix = {arXiv},
       eprint = {astro-ph/0602497},
 primaryClass = {astro-ph},
       adsurl = {https://ui.adsabs.harvard.edu/abs/2006A&A...456..917R}
}

@ARTICLE{Janiuk2022,
       author = {{Janiuk}, Agnieszka and {James}, Bestin},
        title = "{Magnetically arrested accretion disks launching structured jets in application to GRB and AGN engines}",
      journal = {\aap},
         year = 2022,
        month = dec,
       volume = {668},
          eid = {A66},
        pages = {A66},
          doi = {10.1051/0004-6361/202244196},
       adsurl = {https://ui.adsabs.harvard.edu/abs/2022A&A...668A..66J}
}

@ARTICLE{Igumenshchev2008,
       author = {{Igumenshchev}, Igor V.},
        title = "{Magnetically Arrested Disks and the Origin of Poynting Jets: A Numerical Study}",
      journal = {\apj},
         year = 2008,
        month = apr,
       volume = {677},
       number = {1},
        pages = {317-326},
          doi = {10.1086/529025},
archivePrefix = {arXiv},
       eprint = {0711.4391},
 primaryClass = {astro-ph},
       adsurl = {https://ui.adsabs.harvard.edu/abs/2008ApJ...677..317I}
}

@ARTICLE{Narayan2012,
       author = {{Narayan}, Ramesh and {S{\"A} dowski}, Aleksander and {Penna}, Robert F. and {Kulkarni}, Akshay K.},
        title = "{GRMHD simulations of magnetized advection-dominated accretion on a non-spinning black hole: role of outflows}",
      journal = {\mnras},
         year = 2012,
        month = nov,
       volume = {426},
       number = {4},
        pages = {3241-3259},
          doi = {10.1111/j.1365-2966.2012.22002.x},
archivePrefix = {arXiv},
       eprint = {1206.1213},
 primaryClass = {astro-ph.HE},
       adsurl = {https://ui.adsabs.harvard.edu/abs/2012MNRAS.426.3241N}
}

@ARTICLE{Tchekhovskoy2011,
       author = {{Tchekhovskoy}, Alexander and {Narayan}, Ramesh and {McKinney}, Jonathan C.},
        title = "{Efficient generation of jets from magnetically arrested accretion on a rapidly spinning black hole}",
      journal = {\mnras},
         year = 2011,
        month = nov,
       volume = {418},
       number = {1},
        pages = {L79-L83},
          doi = {10.1111/j.1745-3933.2011.01147.x},
archivePrefix = {arXiv},
       eprint = {1108.0412},
 primaryClass = {astro-ph.HE},
       adsurl = {https://ui.adsabs.harvard.edu/abs/2011MNRAS.418L..79T}
}

@ARTICLE{Fender2004a,
       author = {{Fender}, R.~P. and {Belloni}, T.~M. and {Gallo}, E.},
        title = "{Towards a unified model for black hole X-ray binary jets}",
      journal = {\mnras},
         year = 2004,
        month = dec,
       volume = {355},
       number = {4},
        pages = {1105-1118},
          doi = {10.1111/j.1365-2966.2004.08384.x},
archivePrefix = {arXiv},
       eprint = {astro-ph/0409360},
 primaryClass = {astro-ph},
       adsurl = {https://ui.adsabs.harvard.edu/abs/2004MNRAS.355.1105F}
}

@ARTICLE{Meyer-Hofmeister2009,
       author = {{Meyer-Hofmeister}, E. and {Liu}, B.~F. and {Meyer}, F.},
        title = "{The hard to soft spectral transition in LMXBs-affected by recondensation of gas into an inner disk}",
      journal = {\aap},
         year = 2009,
        month = dec,
       volume = {508},
       number = {1},
        pages = {329-337},
          doi = {10.1051/0004-6361/200913044},
archivePrefix = {arXiv},
       eprint = {0910.4273},
 primaryClass = {astro-ph.HE},
       adsurl = {https://ui.adsabs.harvard.edu/abs/2009A&A...508..329M}
}

@ARTICLE{Remillard2022,
       author = {{Remillard}, Ronald A. and {Loewenstein}, Michael and {Steiner}, James F. and {Prigozhin}, Gregory Y. and {LaMarr}, Beverly and {Enoto}, Teruaki and {Gendreau}, Keith C. and {Arzoumanian}, Zaven and {Markwardt}, Craig and {Basak}, Arkadip and {Stevens}, Abigail L. and {Ray}, Paul S. and {Altamirano}, Diego and {Buisson}, Douglas J.~K.},
        title = "{An Empirical Background Model for the NICER X-Ray Timing Instrument}",
      journal = {\aj},
         year = 2022,
        month = mar,
       volume = {163},
       number = {3},
          eid = {130},
        pages = {130},
          doi = {10.3847/1538-3881/ac4ae6},
archivePrefix = {arXiv},
       eprint = {2105.09901},
 primaryClass = {astro-ph.IM},
       adsurl = {https://ui.adsabs.harvard.edu/abs/2022AJ....163..130R}
}

@ARTICLE{Markoff2004,
       author = {{Markoff}, Sera and {Nowak}, Michael A.},
        title = "{Constraining X-Ray Binary Jet Models via Reflection}",
      journal = {\apj},
         year = 2004,
        month = jul,
       volume = {609},
       number = {2},
        pages = {972-976},
          doi = {10.1086/421099},
archivePrefix = {arXiv},
       eprint = {astro-ph/0403468},
 primaryClass = {astro-ph},
       adsurl = {https://ui.adsabs.harvard.edu/abs/2004ApJ...609..972M}
}

@ARTICLE{Markoff2003,
       author = {{Markoff}, Sera and {Nowak}, Michael and {Corbel}, St{\'e}phane and {Fender}, Rob and {Falcke}, Heino},
        title = "{Modeling the X-ray contribution of XRB jets}",
      journal = {\nar},
         year = 2003,
        month = oct,
       volume = {47},
       number = {6-7},
        pages = {491-493},
          doi = {10.1016/S1387-6473(03)00078-2},
archivePrefix = {arXiv},
       eprint = {astro-ph/0301479},
 primaryClass = {astro-ph},
       adsurl = {https://ui.adsabs.harvard.edu/abs/2003NewAR..47..491M}
}

@ARTICLE{Markoff2001,
       author = {{Markoff}, S. and {Falcke}, H. and {Fender}, R.},
        title = "{A jet model for the broadband spectrum of XTE J1118+480. Synchrotron emission from radio to X-rays in the Low/Hard spectral state}",
      journal = {\aap},
         year = 2001,
        month = jun,
       volume = {372},
        pages = {L25-L28},
          doi = {10.1051/0004-6361:20010420},
archivePrefix = {arXiv},
       eprint = {astro-ph/0010560},
 primaryClass = {astro-ph},
       adsurl = {https://ui.adsabs.harvard.edu/abs/2001A&A...372L..25M}
}

@ARTICLE{Malzac2014,
       author = {{Malzac}, Julien},
        title = "{The spectral energy distribution of compact jets powered by internal shocks}",
      journal = {\mnras},
         year = 2014,
        month = sep,
       volume = {443},
       number = {1},
        pages = {299-317},
          doi = {10.1093/mnras/stu1144},
archivePrefix = {arXiv},
       eprint = {1406.2208},
 primaryClass = {astro-ph.HE},
       adsurl = {https://ui.adsabs.harvard.edu/abs/2014MNRAS.443..299M}
}

@ARTICLE{Narayan2003,
       author = {{Narayan}, Ramesh and {Igumenshchev}, Igor V. and {Abramowicz}, Marek A.},
        title = "{Magnetically Arrested Disk: an Energetically Efficient Accretion Flow}",
      journal = {\pasj},
         year = 2003,
        month = dec,
       volume = {55},
        pages = {L69-L72},
          doi = {10.1093/pasj/55.6.L69},
archivePrefix = {arXiv},
       eprint = {astro-ph/0305029},
 primaryClass = {astro-ph},
       adsurl = {https://ui.adsabs.harvard.edu/abs/2003PASJ...55L..69N}
}

@ARTICLE{Makishima1986,
       author = {{Makishima}, K. and {Maejima}, Y. and {Mitsuda}, K. and {Bradt}, H.~V. and {Remillard}, R.~A. and {Tuohy}, I.~R. and {Hoshi}, R. and {Nakagawa}, M.},
        title = "{Simultaneous X-Ray and Optical Observations of GX 339-4 in an X-Ray High State}",
      journal = {\apj},
         year = 1986,
        month = sep,
       volume = {308},
        pages = {635},
          doi = {10.1086/164534},
       adsurl = {https://ui.adsabs.harvard.edu/abs/1986ApJ...308..635M}
}

@article{Ingram_2024,
doi = {10.3847/1538-4357/ad3faf},
url = {https://dx.doi.org/10.3847/1538-4357/ad3faf},
year = {2024},
month = {jun},
publisher = {The American Astronomical Society},
volume = {968},
number = {2},
pages = {76},
author = {Adam Ingram and Niek Bollemeijer and Alexandra Veledina and Michal Dovčiak and Juri Poutanen and Elise Egron and Thomas D. Russell and Sergei A. Trushkin and Michela Negro and Ajay Ratheesh and Fiamma Capitanio and Riley Connors and Joseph Neilsen and Alexander Kraus and Maria Noemi Iacolina and Alberto Pellizzoni and Maura Pilia and Francesco Carotenuto and Giorgio Matt and Guglielmo Mastroserio and Philip Kaaret and Stefano Bianchi and Javier A. García and Matteo Bachetti and Kinwah Wu and Enrico Costa and Melissa Ewing and Vadim Kravtsov and Henric Krawczynski and Vladislav Loktev and Andrea Marinucci and Lorenzo Marra and Romana Mikušincová and Edward Nathan and Maxime Parra and Pierre-Olivier Petrucci and Simona Righini and Paolo Soffitta and James F. Steiner and Jiří Svoboda and Francesco Tombesi and Stefano Tugliani and Francesco Ursini and Yi-Jung Yang and Silvia Zane and Wenda Zhang and Iván Agudo and Lucio A. Antonelli and Luca Baldini and Wayne H. Baumgartner and Ronaldo Bellazzini and Stephen D. Bongiorno and Raffaella Bonino and Alessandro Brez and Niccolò Bucciantini and Simone Castellano and Elisabetta Cavazzuti and Chien-Ting Chen and Stefano Ciprini and Alessandra De Rosa and Ettore Del Monte and Laura Di Gesu and Niccolò Di Lalla and Alessandro Di Marco and Immacolata Donnarumma and Victor Doroshenko and Steven R. Ehlert and Teruaki Enoto and Yuri Evangelista and Sergio Fabiani and Riccardo Ferrazzoli and Shuichi Gunji and Kiyoshi Hayashida and Jeremy Heyl and Wataru Iwakiri and Svetlana G. Jorstad and Vladimir Karas and Fabian Kislat and Takao Kitaguchi and Jeffery J. Kolodziejczak and Fabio La Monaca and Luca Latronico and Ioannis Liodakis and Simone Maldera and Alberto Manfreda and Frédéric Marin and Alan P. Marscher and Herman L. Marshall and Francesco Massaro and Ikuyuki Mitsuishi and Tsunefumi Mizuno and Fabio Muleri and Chi-Yung Ng and Stephen L. O’Dell and Nicola Omodei and Chiara Oppedisano and Alessandro Papitto and George G. Pavlov and Abel L. Peirson and Matteo Perri and Melissa Pesce-Rollins and Andrea Possenti and Simonetta Puccetti and Brian D. Ramsey and John Rankin and Oliver J. Roberts and Roger W. Romani and Carmelo Sgrò and Patrick Slane and Gloria Spandre and Douglas A. Swartz and Toru Tamagawa and Fabrizio Tavecchio and Roberto Taverna and Yuzuru Tawara and Allyn F. Tennant and Nicholas E. Thomas and Alessio Trois and Sergey S. Tsygankov and Roberto Turolla and Jacco Vink and Martin C. Weisskopf and Fei Xie and (IXPE Collaboration)},
title = {Tracking the X-Ray Polarization of the Black Hole Transient Swift J1727.8–1613 during a State Transition},
journal = {The Astrophysical Journal}
}
\bibliographystyle{aasjournal}
\appendix

\begin{center}
\begin{longtable}{cccccll}
\caption{Details of the observations of the three sources used in this study, as well as corona heights obtained from our X-ray spectral modeling.   }\\
\multicolumn{7}{c}{MAXI\,J1348--630}                                                                                                                                                                               \\
Epoch                                     & \multicolumn{1}{l}{Date (MJD)} & \multicolumn{1}{l}{State} & \multicolumn{1}{l}{NuSTAR ObsID} & \multicolumn{1}{l}{Supp. Data \footnote{Ni=\textit{NICER}, S = \textit{Swift}/XRT}} & Supp. ObsID & Corona Heights ($R_g$)\footnote{\textit{p} means model pegged height at minimum value,-p means lower limit is minimum value} \\ \hline
1                                         & 58515.13                       & LHS                       & 80402315002                      & -                              &             & $2.11\pm 0.04$ \\
2                                         & 58515.60                       & LHS                       & 80402315004                      & -                              &             & $2.13^{+0.05}_{-0.04}$ \\
3                                         & 58521.07                       & HIMS                      & 80402315006                      & S                              & 00011107004 & $5.2^{+0.4}_{-0.3}$    \\
4                                         & 58525.26                       & SIMS                      & 80402315008                      & -                              &             & $2.5\pm0.1$ \\
5                                         & 58550.64                       & HSS                       & 80402315010                      & S                              & 00088843002 & $2.01^{p}$             \\
\end{longtable}
\label{htable}
\end{center}

\clearpage
\movetableright=-1in
\begin{table}
\footnotesize
\setlength{\tabcolsep}{3pt}
\begin{tabular}{lllllllllll}
\multicolumn{11}{c}{MAXI\,J1348--630}                                                                                                                                                                                                                                                                                                               \\
Model                               & Parameter      & Epoch 1                       & Epoch 2                      & Epoch 3                  & Epoch 4                      & Epoch  5                         & Epoch 6                     & Epoch 7                       & Epoch 8                    & Epoch 9                       \\
\hline
\texttt{tbabs}                     & nH\,($\times10^{22}$\,atoms\,cm$^{-2}$)\footnote[5]{nH, a, i, and A$_{\mathrm{Fe}}$ tied across all epochs in each group}             & $0.91 \pm 0.01$               &                              &                          &                              &                               &                             &                               &                            &                               \\
\texttt{diskbb}                                    & T$_{\mathrm{in}}$\,(keV)       & $0.49^{+0.02}_{-0.01}$        & $0.54 \pm 0.02$              & $0.69 \pm 0.01$          & $0.78 \pm 0.01$              & $0.55 \pm 0.01$               & $0.55 \pm 0.01$             & $0.26 \pm 0.01$               & $1.0 \pm 0.03$             & $0.20\pm 0.01$     \\
                                    & N$_{\mathrm{d}}$        & $1.6e4 \pm 3e3$               & $9.1e3 \pm 2e3$            & $2.0e4\pm 1e3$ & $2.1e4 \pm 1e3$              & $1.0e5 \pm 1e4$               & $3.7e4 \pm 5e3$             & $3.9e4^{+5e3}_{-4e3}$         & $51^{+4}_{-6}$             & $6.5e4\pm 7e3$ \\
\hline
\end{tabular}
\label{m1348_table}
\end{table}

\clearpage

\movetableright=-1in
\begin{table}
\footnotesize
\setlength{\tabcolsep}{3pt}
\begin{tabular}{llllllllllll}
\multicolumn{10}{c}{MAXI\,J1820+070 Group 1}                                                                                                                                                                                                         &  &  \\
Model                               & Par      & Epoch 1             & Epoch 2               & Epoch 3            & Epoch 4                   & Epoch 5              & Epoch 6               & Epoch 7           & Epoch 8                    &  &  \\
\hline
\texttt{tbabs}                     & nH\,($\times10^{22}$\,atoms\,cm$^{-2}$)             & $0.39\pm0.01$     &                       &                    &                           &                      &                       &                   &                            &  &  \\
\texttt{diskbb}                                    & T$_{\mathrm{in}}$\,(keV)           & $0.52 \pm 0.01$     & $0.58 \pm 0.01$       & $0.19 \pm 0.01$    & $0.193 \pm 0.001$         & $0.192\pm 0.001$     & $0.190 \pm 0.001$      & $0.189 \pm 0.001$ & $0.190 \pm 0.001$         &  &  \\
                                    & N$_{\mathrm{d}}$          & $1310^{+55}_{-29}$  & $7400^{+400}_{-300}$  & $2.7e6\pm 1e5$     & $2.3e6^{+8e5}_{-6e4}$     & $2.4e6^{+6e4}_{-4e4}$& $2.2e6^{+5e4}_{-3e4}$ & $2.3e6 \pm 1e5$   & $2.1e6 \pm 1e4$       &  &  \\
\texttt{bbody}                      & kT\,(keV)             & $1.18 \pm 0.03$     & $1.30^{+0.01}_{-0.02}$& $0.43 \pm 0.01$    & $0.47 \pm 0.01$           & $0.46 \pm 0.01$      & $0.43 \pm 0.01$       & $0.42 \pm 0.01$   & $0.43 \pm 0.01$           &  &  \\
                                    & N$_{\mathrm{bb}}$           & $< 0.01$          & $0.079 \pm 0.001$     & $0.043 \pm 0.001$  & $0.045 \pm 0.001$         & $0.045 \pm 0.001$    & $0.040 \pm 0.001$     & $0.042 \pm 0.001$ & $0.04 \pm 0.001$            &  &  \\
\hline
\end{tabular}
\end{table}
\label{m1820_1_table}

\vspace{5cm}

\movetableright=-1in

\begin{table}
\footnotesize
\setlength{\tabcolsep}{3pt}
\begin{tabular}{llllllllllll}
\multicolumn{10}{c}{MAXI\,J1820+070 Group 1 with Hydrogen Column Density Fixed}                                                                                                                                                                                                         &  &  \\
Model                               & Par       & Epoch 1               & Epoch 2               & Epoch 3            & Epoch 4                   & Epoch 5              & Epoch 6               & Epoch 7           & Epoch 8                    &  &  \\
\hline
\texttt{diskbb}                     & nH ($\times10^{22}$\,atoms\,cm$^{-2}$)        & $0.07^{\mathrm{fixed}}$            &                       &                    &                           &                      &                       &                   &                            &  &  \\
                                    & T$_{\mathrm{in}}$\,(keV)  & $0.49 \pm 0.01$       & $0.55 \pm 0.01$       & $0.29 \pm 0.01$    & $0.308 \pm 0.003$         & $0.300\pm 0.001$     & $0.300 \pm 0.002$     & $0.314 \pm 0.003$ & $0.318 \pm 0.002$         &  &  \\
                                    & N$_{d}$   & $1722^{+58}_{-62}$    & $9600^{+300}_{-500}$  & $8.1e4\pm 2e3$     & $7.2e4 \pm 3e3 $          & $7.5e4 \pm 3e3$       & $5.8e4\pm 2e3 $       & $4.3e4 \pm 2e3$   & $4.8e4 \pm 2e3$       &  &  \\
\texttt{bbody}                      & kT\,(keV)        & $1.13 \pm 0.03$       & $1.25\pm 0.02$        & $0.61 \pm 0.01$    & $0.70 \pm 0.01$           & $0.64 \pm 0.01$      & $0.65 \pm 0.01$       & $0.58 \pm 0.02$   & $0.66 \pm 0.01$           &  &  \\
                                    & N$_{bb}$  & $< 0.01$              & $0.079 \pm 0.001$     & $0.025 \pm 0.001$  & $0.036 \pm 0.001$         & $0.027 \pm 0.001$    & $0.023 \pm 0.001$     & $0.014 \pm 0.001$ & $0.023 \pm 0.001$            &  &  \\
\hline
\end{tabular}
\end{table}
\label{m1820_1_table_nhfix}

\clearpage

\movetableright=-1.0in
\begin{table}
\scriptsize
\setlength{\tabcolsep}{2pt}
\begin{tabular}{llllllllllllll}
\multicolumn{11}{c}{MAXI\,J1820+070 Group 2} \\
Model                               & Par      & Epoch 9               & Epoch 10 \footnote{Disk not detected}                   & Epoch 11 \footnote{Disk and blackbody component not detected}                 & Epoch 12                      & Epoch 13                      & Epoch 14                    & Epoch 15                    & Epoch 16                      & Epoch 17                 \\
\hline
\texttt{tbabs}                     & nH\,($\times10^{22}$\,atoms\,cm$^{-2}$)\,             & $0.072 \pm 0.001$     &                            &                           &                               &                               &                             &                             &                               &                          \\
\texttt{diskbb}                                    & T$_{\mathrm{in}}$\,(keV)            & $0.30 \pm 0.01$       &                            &                           & $0.28\pm0.01$     	      & $0.29 \pm 0.01$               & $0.688 \pm 0.003$           & $0.65 \pm 0.01$             & $0.68\pm0.01$                 & $0.69\pm0.01$             \\
                                    & N$_{\mathrm{d}}$\,($\times10^{4}$)         & $5.5 \pm 0.3 $       &                            &                           & $8.0^{+0.6}_{-0.5}$         & $4.7^{+0.3}_{-0.4}$         & $2.04^{+0.04}_{-0.03}$       & $2.83\pm0.07$               & $2.3^{+0.1}_{-0.2}$     & $2.09^{+0.07}_{-0.05}$  \\
\texttt{bbody}                      & kT\,(keV)             & $0.54 \pm 0.02$       & $0.60 \pm 0.04$            &                           & $0.52\pm 0.02$                & $0.56 \pm 0.02$               & $1.17 \pm 0.01$             & $0.80\pm 0.01$              & $0.89\pm0.02$                 & $0.88\pm0.01$             \\
\hline
\end{tabular}
\label{m1820_2_table}
\end{table}

\clearpage
\begin{tabular}{llllllll}
\multicolumn{8}{c}{MAXI\,J1820+070 Group 3} \\                                                                                                                                                  \\
Model                               & Par            & Epoch 18                & Epoch 19                & Epoch 20                   & Epoch 21                    & Epoch 22                    & Epoch 23                    \\
\hline
\texttt{tbabs}                     & nH\,($\times10^{22}$\,atoms\,cm$^{-2}$)             & $0.07 \pm 0.01$         &                         &                            &                             &                             &                             \\
\texttt{diskbb}                                    & T$_{\mathrm{in}}$\,(keV)       & $0.67 \pm 0.01$         & $0.578 \pm 0.02$        & $0.27\pm0.01$              & $0.51\pm 0.03$              & $1.0 \pm 0.3$             & $1.0\pm0.1$       \\
                                    & N$_{\mathrm{d}}$        & $1.8e4^{+7e2}_{-1e3}$   & $2.8e4^{+3.3e3}_{-5e2}$ & $1.1e5^{+2e4}_{-1e4}$      & $360 \pm 130$               & $0.6\pm 0.5$                & $0.8 \pm 0.3$   \\
\texttt{bbody}                      & kT\,(keV)             & $0.90 \pm 0.02$         & $0.70^{+0.01}_{-0.03}$  & $0.46 \pm 0.02$            & $1.17^{+0.08}_{-0.13}$      &                             &                             \\
                                    & N$_{\mathrm{bb}}$       & $0.04 \pm 0.01$         & $0.11^{+0.03}_{-0.01}$  & $0.019 \pm 0.002$          & $<0.001$       &                                          &                             \\
\hline
\end{tabular}
\label{m1820_3_table}

\clearpage

\begin{table}
\footnotesize
\setlength{\tabcolsep}{3pt}
\begin{tabular}{llllllllllll}
\multicolumn{10}{c}{MAXI\,J1535--571 Group 1}                                                                                                                                                                                                                                                                                                  \\
Model                               & Par      & Epoch 1                          & Epoch 2                                   & Epoch 3                                & Epoch 4                                   & Epoch  5                                   & Epoch 6                                      & Epoch 7                     & Epoch 8                     \\
\hline
\texttt{tbabs}                     & nH\,($\times10^{22}$\,atoms\,cm$^{-2}$)            & $4.07\pm 0.02$         &                                           &                                        &                                           &                                            &                                              &                             &                             \\
\texttt{diskbbb}                                    & T$_{\mathrm{in}}$\,(keV)      & $0.37\pm 0.01$         & $0.128 \pm 0.002$                         & $0.168^{+0.001}_{-0.003}$              & $0.213^{+0.006}_{-0.002}$                 & $0.116^{+0.001}_{-0.002}$                  & $0.109 \pm 0.002$                            & $1.25\pm 0.01$              & $1.23\pm 0.01$      \\
                                    & N$_{\mathrm{d}}$       & $1.2e5\pm2e4$          & $1.9e8\pm3e7$                             & $2.3e7^{+3e6}_{-2e6}$                  & $1.3e8^{+3e7}_{-4e7}$                     & $6.7e8^{+1.1e8}_{-8e7}$                    & $1.3e9\pm2e8$                                & $1454^{+77}_{-47}$          & $1897^{+53}_{-49}$        \\
\hline
\end{tabular}
\label{m1535_1_table}
\end{table}

\clearpage
\begin{table}
\footnotesize
\setlength{\tabcolsep}{3pt}
\begin{tabular}{llllllllllll}
\multicolumn{10}{c}{MAXI\,J1535--571 Group 2}                                                                                                                                                                                                                             \\
Model                               & Par            & Epoch 9                   & Epoch 10                    & Epoch 11                  & Epoch 12                            & Epoch  13                           & Epoch 14                     & Epoch 15                  & Epoch 16                   \\
\hline
\texttt{tbabs}                     & nH\,($\times10^{22}$\,atoms\,cm$^{-2}$)             & $4.11^{+0.10}_{-0.05}$    &                             &                           &                                     &                                     &                              &                           &                            \\
\texttt{diskbb}                                    & T$_{\mathrm{in}}$\,(keV)       & $1.14 \pm 0.01$           & $1.15 \pm 0.01 $            & $1.14\pm 0.01$            & $0.30\pm 0.01$                      & $0.28\pm 0.01$                      & $0.57\pm 0.03$               & $0.818 \pm 0.003$         & $0.82 \pm 0.01$  \\
                                    & N$_{\mathrm{d}}$        & $4506^{+105}_{-52}$       & $1659^{+54}_{-62}$          & $2198^{+55}_{-36}$        & $1.5e6\pm 3e5$                      & $3.2e6^{+8e5}_{-5e5}$               & $5200\pm 1300$               & $5110^{+160}_{-90}$       & $4865^{+180}_{-140}$ \\
\texttt{relxilllp}                  & h\,(r$_\mathrm{g}$)              & $2.01^{p}$               & $2.03^{+0.05}_{-p}$         & $2.04^{+0.06}_{-p}$       & $2.9^{+0.2}_{-0.1}$                 & $2.6\pm 0.1$                        & $2.3 \pm 0.1$                & $2.01^{p}$               & $2.01^{p}$      \\
\end{tabular}
\end{table}

\begin{figure}[ht!]
\centering
\includegraphics[width=15cm]{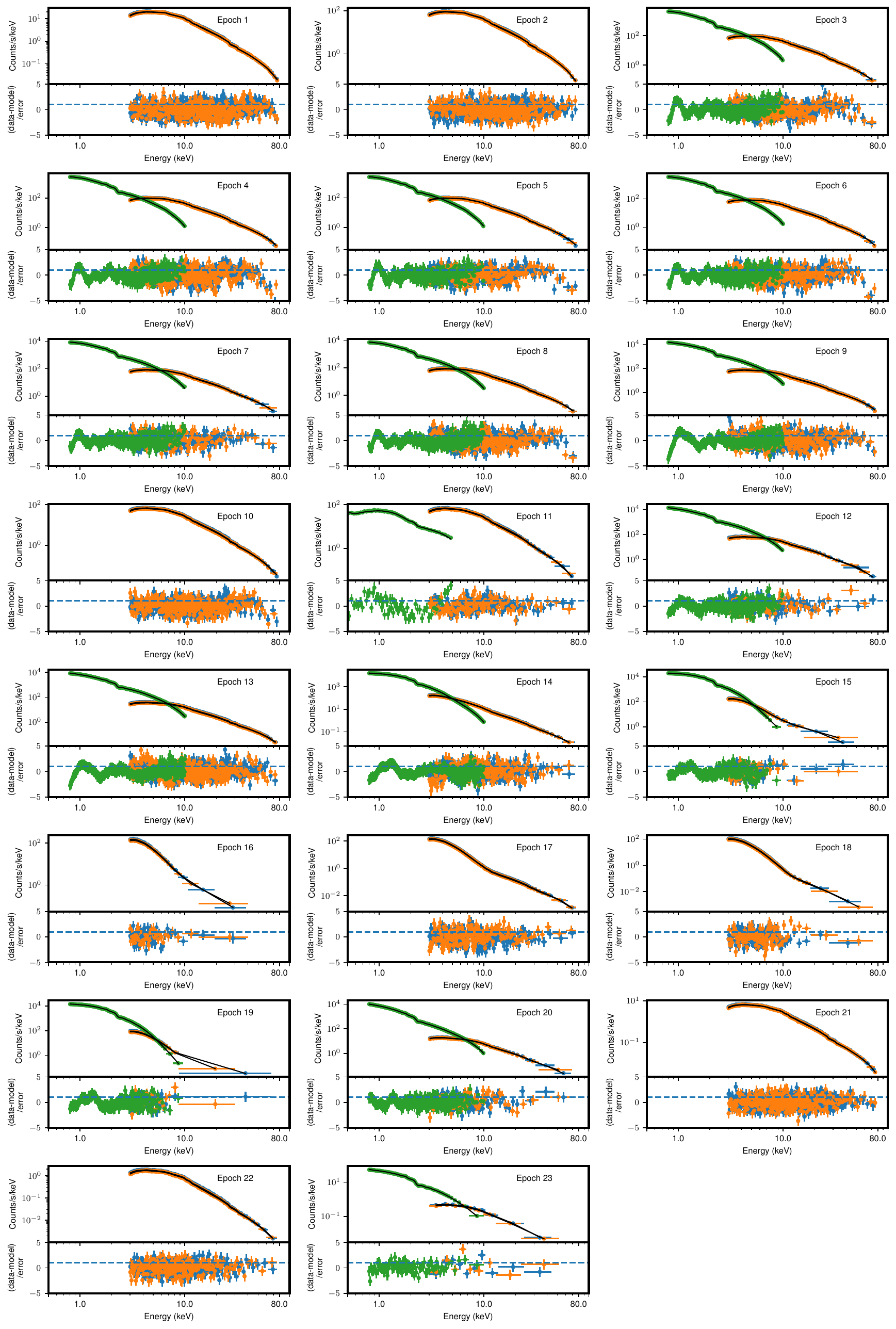}
\caption{The data and spectral model comprising each epoch of MAXI\,J1820+070's 2018 outburst (top panels). Bottom panels show (Data-Model)/Error. Plots are rebinned for illustrative purposes. 
\label{1820_grid}}
\end{figure}
\clearpage
\begin{figure}[ht!]
\centering
\includegraphics[width=15cm]{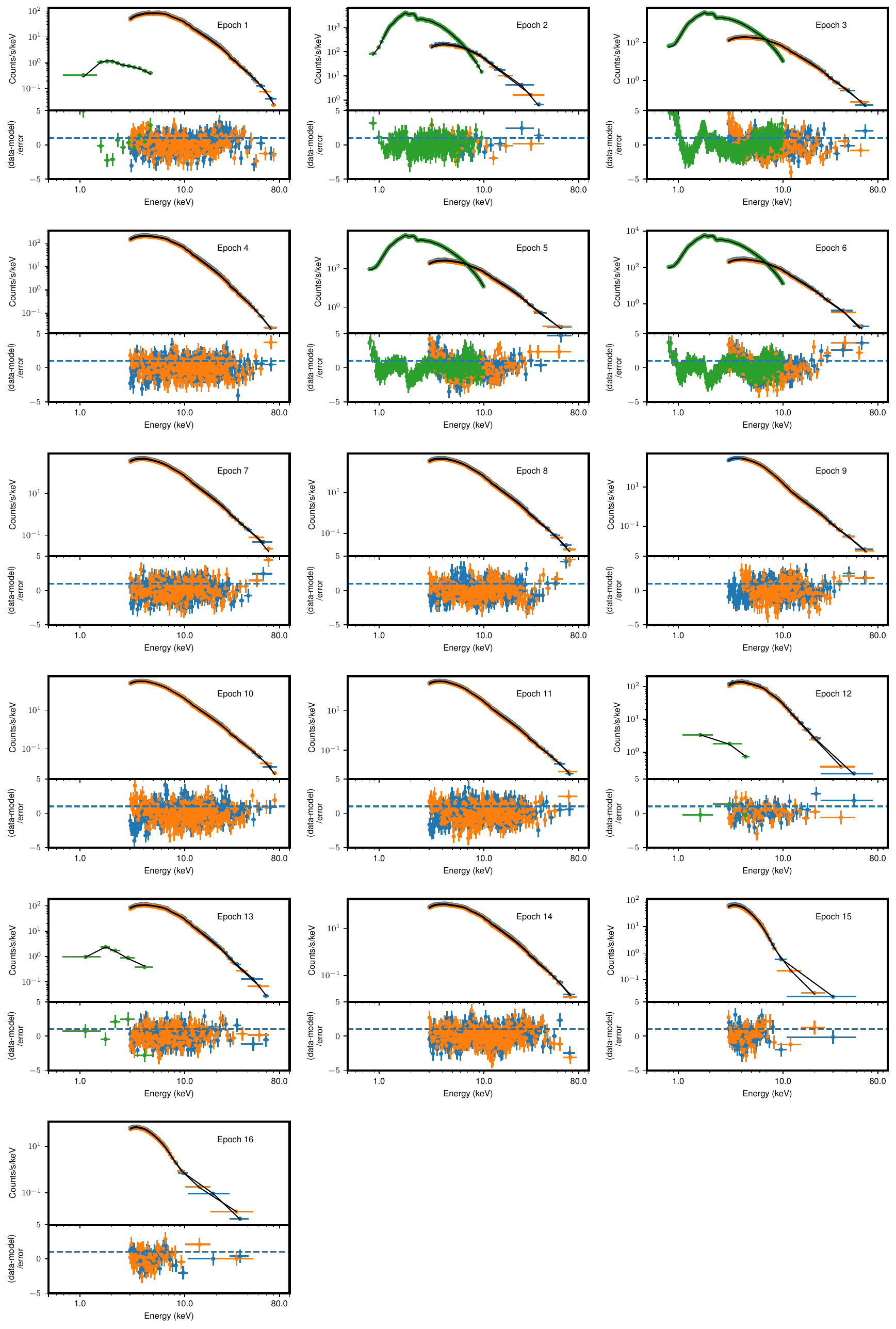}
\caption{The data and spectral model comprising each epoch of  MAXI\,J1535\--571 2017 outburst (top panels). Bottom panels show (Data-Model)/Error. Plots are rebinned for illustrative purposes. 
\label{1535_grid}}
\end{figure}
\clearpage
\begin{figure}[ht!]
\centering
\includegraphics[width=17cm]{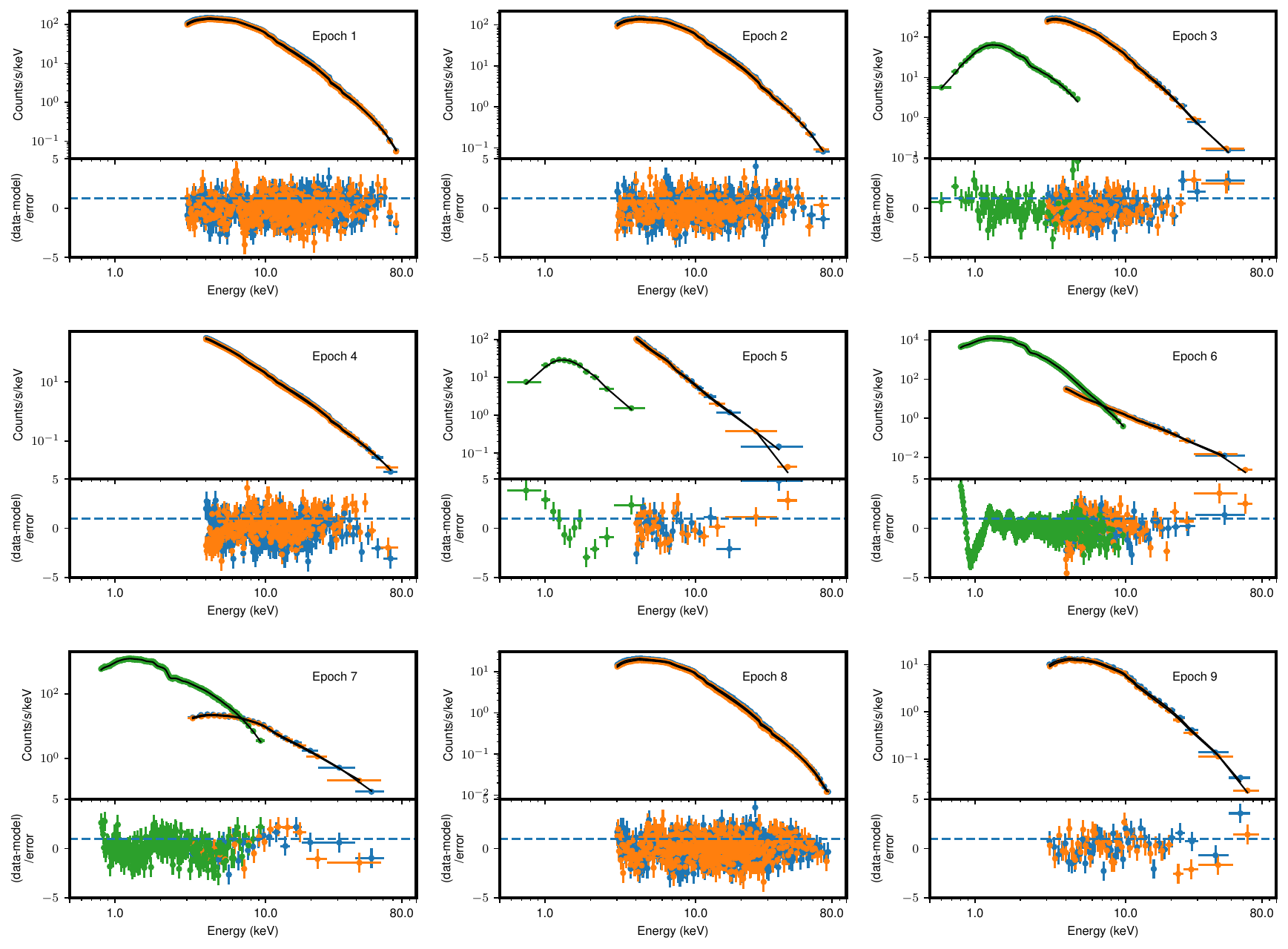}
\caption{The data and spectral model comprising each epoch of  MAXI\,J1348\--630 2019 outburst (top panels). Bottom panels show (Data-Model)/Error.  Plots are rebinned for illustrative purposes. 
\label{1348_grid}}
\end{figure}

\end{document}